\documentclass[final,3p,times]{elsarticle}

\usepackage{amsthm}

\usepackage{amsmath}

\usepackage{stmaryrd}

\usepackage{color}
\usepackage[table, dvipsnames, svgnames]{xcolor}

\usepackage{float}

\usepackage{framed}

\usepackage[shortlabels]{enumitem}

\usepackage{graphicx, makecell}
\usepackage{multirow}

\usepackage{algorithm}
\usepackage{algpseudocode}

\usepackage{tikz}
\usetikzlibrary{shapes.geometric,arrows,positioning,fit,calc}

\usepackage{subcaption}

\definecolor{light-light-gray}{gray}{0.90}
\tikzstyle{rectangle-b} = [rectangle, rounded corners, minimum width=3cm, minimum height=1cm, text centered, font=\normalsize, color=black, draw=black, line width=2, fill=White]
\tikzstyle{rectangle-v} = [rectangle, minimum width=7cm, minimum height=3.2cm, text centered, font=\normalsize, color=black, draw=black, line width=2, fill=light-light-gray]
\tikzstyle{rectangle-o} = [rectangle, minimum width=3cm, minimum height=1cm, text centered, font=\normalsize, color=black, draw=black, line width=2, fill=White]
\tikzstyle{rectangle-g} = [rectangle, minimum width=3cm, minimum height=1cm, text centered, font=\normalsize, color=black, draw=black, line width=2, fill=White]
\tikzstyle{large-rectangle-o} = [rectangle, minimum width=7cm, minimum height=1cm, text centered, font=\normalsize, color=black, draw=black, line width=2, fill=White]
\tikzstyle{rectangle-h} = [rectangle, minimum width=3.5cm, minimum height=4.4cm, text centered, font=\normalsize, color=black, draw=black, dashed, line width=2, fill=White]
\tikzstyle{diamond-g} = [diamond, aspect=2, minimum width=3cm, minimum height=1cm, text centered, font=\normalsize, color=black, draw=black, line width=2, fill=White]
\tikzstyle{arrow} = [thick, draw=black, line width=2, ->, >=stealth]
\tikzstyle{-arrow} = [thick, draw=black, line width=2, <-, >=stealth]

\usepackage{enumitem,amssymb}
\newlist{todolist}{itemize}{2}
\setlist[todolist]{label=$\square$}

\usepackage[colorlinks]{hyperref}

\usepackage[textsize=small]{todonotes}

\journal{any Elsevier journal}

\AtBeginDocument{%
  \hypersetup{
    citecolor=cyan,
    linkcolor=cyan,   
    urlcolor=cyan}}

\begin{document}

\begin{frontmatter}



\title{Bayesian and non-Bayesian multi-fidelity surrogate models for multi-objective aerodynamic optimization under extreme cost imbalance}

\author[SU]{Marc Schouler}
\author[SU]{Anca Belme}
\author[SU]{Paola Cinnella}

\affiliation[SU]{organization={Institut Jean Le Rond D'Alembert, Sorbonne University, CNRS},
            addressline={4 Place Jussieu}, 
            city={Paris},
            postcode={75005}, 
            country={France}}

\begin{abstract}
Aerodynamic shape optimization in industry still faces challenges related to robustness and scalability. This aspect becomes crucial for advanced optimizations that rely on expensive high-fidelity flow solvers, where computational budget constraints only allow a very limited number of simulations within the optimization loop. To address these challenges, we investigate strategies based on multi-fidelity surrogate models. In particular, we focus on the case of extreme computational cost imbalance between the high- and low-fidelity models, which severely limits the maximum allowable number of high-fidelity function calls. To maximize the information extracted from the high-fidelity samples, we generate a reduced representation of the design space and use an adaptive infill strategy to smartly place the high-fidelity samples where they can best guide the optimization. Bayesian co-kriging and non-Bayesian multi-fidelity neural networks are trained by combining low- and high-fidelity models for a use-case consisting of a low Reynolds linear outlet guide vane at subsonic and transitional flow conditions. Coarse-mesh RANS simulations are used as low-fidelity model while RANS simulations with a transition model and automatically (feature-based) adapted meshes are chosen as the high-fidelity one. Each surrogate model is then associated to an infill strategy of its kind and a proper orthogonal decomposition of the shape parametrization is used to reduce by half the dimension of the problem. Based on inverted distance and hypervolume metrics, we find that the simpler co-kriging representation in conjunction with Bayesian infill yields better performance than the multi-fidelity neural network and the considered non-Bayesian method.
\end{abstract}

\begin{graphicalabstract}
\end{graphicalabstract}

\begin{highlights}
\item An aerodynamic shape optimization problem is solved with multi-fidelity deep neural network and Co-kriging.
\item Bayesian and non-Bayesian infilling strategies are compared considering each model's paradigm.
\item Design parameter reduction is used to decrease the problem dimension.
\item Robustness to deformation is ensured by using mesh adapted high-fidelity solutions.
\end{highlights}


\end{frontmatter}


\section{Introduction}\label{sec:introduction}
As key component of aviation design, aerodynamic shape optimization (ASO) has for long been a major challenge of the aeronautics industry. With the current need to address environmental sustainability concerns as well as the competitiveness of the sector, it counts among the principal means to improve aircraft performances \cite{le_clainche_improving_2023,afonso_strategies_2023}.
In the frame of this paper, ASO designates engineering methods that are used to improve the aerodynamic properties of external components \cite{skinner_state---art_2018}, or the efficiency of turbomachinery systems \cite{li_review_2017}, through parametrized shape deformation. \par

ASO methods are usually categorized as gradient-based and gradient-free methods. For the former category, the optimal solution is sought by computing the objective function derivative with respect to the problem parameters usually using the adjoint method \cite{nadarajah2007optimum}. For the latter, which mostly consists in genetic or particle swarm algorithms \cite{ampellio2016turbomachinery,karbasian2022gradient}, the optimal solution is sought by repetitive calls to the objective function for evolving generations (or populations) of candidates. The main advantage of both kinds of techniques is their capacity to deal with high-dimensional design spaces but not at the same cost. Indeed, the required number of function evaluations to converge varies with a linear trend for gradient-based optimization while the tendency is quadratic or even cubic for gradient-free methods \cite{li_machine_2022}. Nevertheless, gradient features implementations are still scarce among CFD solvers and gradient-free methods offer several advantages such as an easy way to deal with multiple objectives, multi-modal problems and complex constraints. They are therefore more popular, despite being more costly because of the large number of required function calls. \par

Even though ASO methods in general have greatly benefited from the development of computational fluid dynamics (CFD) and the progress in high-performance computing, their use in industry still poses several difficulties related in particular to the geometry parametrization, the scalability with respect to the number of dimensions, and the preservation of solution accuracy under large shape deformations and modifications of the initial computational grid \cite{martins_aerodynamic_2022}. 
To mitigate the exponential growth of the computational cost incurred under evolutionary algorithms in high-dimensional parameter spaces, surrogate models \cite{forrester2008engineering,forrester2009recent} are used to replace the costly CFD evaluations of the cost functions.  Even then, surrogates are subject to the so called "curse of dimensionality", and the number of CFD solutions required to train them quickly escalates with the dimension of the design space, leading to prohibitive computational costs. This can be alleviated by using multi-fidelity surrogates (e.g. \cite{le_gratiet_recursive_2014,brevault_overview_2020}, see also \cite{fernandez2023review} for a review), which leverage cost function evaluations based on both a low-accuracy but inexpensive model and a costly high-fidelity model to train the surrogate. In addition, adaptive infill techniques have been proposed in the literature to progressively enrich the surrogate model with new samples as the optimization proceeds \cite{keane_statistical_2012,fuhg2021state}. \par

In multi-fidelity ASO however, the fidelity hierarchy often consists of using different levels of mesh refinement \cite{zhang_multi-fidelity_2021}, or of using high-fidelity models allowing for a sufficiently large number of function evaluations \cite{wu_multi-fidelity_2024}. In turns, the cost ratio between the expensive high-fidelity and the cheap low-fidelity computations is of about 10 to 100 and the infill criteria may take into account the fidelity cost to decide with which fidelity the infill samples should be computed \cite{charayron_towards_2023, mourousias_novel_2024}. However, in cases of extreme cost imbalance — such as when expensive unsteady RANS \cite{leusink2015multi} or, even more so, large-eddy simulations (LES) \cite{matar_camille_analysis_2024, matar2025cost} are used as the high-fidelity source — the cost ratio typically falls between $10^3$ and $10^4$ for Reynolds numbers on the order of $10^5$ and $10^6$. In this case, the decision of when to compute high-fidelity infill must be taken explicitly beforehand and with extreme care as the computational budget only allows for a few well-targeted shots to be taken \cite{matar_camille_analysis_2024, matar2025cost}. \par

On the other hand, the high number of function calls may limit the use of fine computational grids used in the CFD simulations. The resulting numerical errors affect the optimization results both in terms of accuracy and convergence speed \cite{cinnella_convergence_2013}. Even using well-converged fine grids for the initial geometry, morphing or remeshing needed to adapt the mesh to the new geometries does not guarantee error-free solutions, which also affects the quality of the surrogate model.  Automatic mesh adaptation \cite{alauzet_decade_2016} is an attractive methodology to control numerical errors by automatically adjusting the computational grid to meet some error criterion. However, only a few authors have employed it in shape optimization \cite{chen_output-based_2019,chen_variable-fidelity_2020,john_using_2020}. \par

In this study, we introduce and compare various multi-fidelity optimization strategies under the assumption of a very high cost imbalance between the levels of fidelity, and hence of a strong constraint on the maximum number of high-fidelity samples. For that purpose, we focus on a use case that allows the computation of a brute-force reference solution while still being representative of a realistic flow configuration, namely, a linear outlet guide vane (OGV) cascade. We develop multi-fidelity surrogates where the fidelity levels differ both in terms of mesh resolution and of RANS turbulence models. To ensure high-quality solutions for the high-fidelity model, we use feature-based anisotropic mesh adaptation \cite{alauzet_feature-based_2021}.  More specifically, we suggest and compare two radically different optimization and surrogate modeling strategies. On one hand, we consider a Bayesian strategy relying on a stochastic Gaussian process surrogate and uncertainty estimates to guide the optimization. On the other hand, we evaluate a multi-objective genetic algorithm assisted by a multi-fidelity deep-learning surrogate. The two approaches are compared under the lens of strong computational budget constraints. %
The main contributions of this paper are summarized below:
\begin{enumerate}[1)]
    \item Multi-fidelity optimization results obtained with both multi-fidelity deep neural networks and co-kriging models are compared for a turbomachinery multi-objective ASO problem.
    \item Dimensionality reduction and adaptive infills are used to improve the surrogate accuracy under a very small number of high-fidelity samples.
    \item To ensure numerical error-free high-fidelity solutions, we use a feature-based mesh adaptation algorithm to generate the high-fidelity samples.
\end{enumerate}
\par

The paper is structured as follows. \autoref{sec:sota} reviews state-of-the-art practices in ASO relative to the above-mentioned contributions. The present Bayesian and non-Bayesian methodologies are then introduced in \autoref{sec:methodology}. Preliminary tests of the surrogate models and optimization strategies using analytical functions are performed in \autoref{sec:validation}. Finally, \autoref{sec:cascade} focuses on the low Reynolds number (LRN) OGV cascade multi-objective optimization problem. The optimization objectives, simulation results for the baseline and a reference optimization are first presented. Then, the results of multiple optimization configurations are compared. Conclusions and perspectives for future investigations are discussed in \autoref{sec:concl}.

\section{State of the art of ASO practices}\label{sec:sota}
\subsection{Multi-fidelity surrogate assisted optimization}
The use of surrogate models of the objective functions to replace expensive CFD solvers in gradient-free optimization has become a popular practice \cite{li_machine_2022}. Furthermore, in response to the hierarchical fidelity and cost of CFD solvers in terms of model or mesh refinement levels, there has been an increasing focus on multi-fidelity surrogate models \cite{li_multi-fidelity_2024}. Among them, the co-kriging method (AR1), based on Kennedy and O'Hagan linear autoregressive information fusion scheme \cite{kennedy_predicting_2000} is by far the  most commonly used in engineering design  \cite{brevault_overview_2020}. The approach has been extended to non-linearly correlated low- and high-fidelity sources by Perdikaris et al. \cite{perdikaris_nonlinear_2017}, who have proposed the non-linear auto-regressive multi-fidelity Gaussian process (NARGP) model. More recently, deep learning based multi-fidelity methods were introduced. Examples include multi-fidelity deep neural networks (MFDNN) \cite{meng_composite_2020}, multi-fidelity deep Bayesian neural networks \cite{meng_multi-fidelity_2021}, multi-fidelity deep Gaussian process \cite{cutajar_deep_2019}, multi-fidelity deep neural operators \cite{howard_multifidelity_2023} and multi-fidelity reinforcement learning \cite{bhola_multi-fidelity_2023}. Multi-fidelity surrogate models for optimization are reviewed in \cite{li_multi-fidelity_2024}. An examination of the number of publications for different multi-fidelity surrogate methods applied to engineering design and optimization shows that correction-based methods, which use additive and/or multiplicative correction terms to model the relationship between successive fidelities, are the most popular (39.1\%), followed by AR1 methods (20.7\%) and single models\footnote{The single model approach consists of using independent surrogate models for each fidelity without assuming any kind of correlation between them.} (17.2\%). Of course, newer methods such as non-linear auto-regressive Gaussian process (2.3\%) or deep Gaussian processes (3.5\%) have been used to a lesser extent. 
In addition, multi-fidelity surrogates under severe computational budget constraints for the high-fidelity model have received little attention in the literature, and the question of whether newer models (theoretically capable of learning any kind of low-/high-fidelity correlation) should be prioritized over simpler models is still open.

\subsection{Adaptive infill}
Multi-fidelity surrogates are generally trained using a reduced number of samples, especially in the high-fidelity level, to reduce the computational cost of generating the training data. However, their accuracy may not be sufficient to guide the optimization algorithm towards an optimum of the actual objective function. To remedy this limitation, the model can be adaptively retrained several times during the optimization by carefully selecting additional samples that are likely to improve the model accuracy in the region of interest. Such a procedure is called adaptive infill or adaptive sampling,  and constitutes an active field of research, especially for multi-fidelity optimization \cite{zhang_multi-fidelity_2021, wu_multi-fidelity_2024, charayron_towards_2023, he_efficient_2024, mourousias_novel_2024}.\par

\autoref{tab:opt} summarizes the main characteristics of recent multi-fidelity optimization strategies with adaptive infill. When surrogate models are of Bayesian nature, Bayesian acquisition functions can be used to select the next infill locations based on the current state of the surrogate model. This paradigm is called Bayesian Optimization (BO), its multi-fidelity multi-objective version being referred-to as MF-MOBO \cite{charayron_towards_2023, he_efficient_2024, mourousias_novel_2024}. Multi-objective acquisition functions mostly consist in generalizations of their single-fidelity equivalent.  For instance the generalized expected improvement matrix (GEIM), also called variable fidelity expected improvement matrix (VFEIM), is an extension of the well known expected improvement (EI) that can be used to derive multi-objective infill criteria \cite{zhan_expected_2017, he_variable-fidelity_2022}. Similarly, the probability of improvement (PI) has been extended to multi-objective adaptive sampling in \cite{keane_statistical_2012} and as the minimal probability of improvement (MPI) criterion in \cite{rahat_alternative_2017}. For non-Bayesian models, the alternative is to combine the surrogate model with an optimization algorithm from which one or multiple best-fit candidates are selected as infill samples to be computed with the high-fidelity solver, while another strategy is used to sample low-fidelity infill samples. In \cite{zhang_multi-fidelity_2021}, the particle swarm optimization (PSO) algorithm is used to sample high-fidelity infills and the max-min Euclidean distance (ED) criterion is used to sample additional low-fidelity infills. In \cite{wu_multi-fidelity_2024}, the teaching-learning-based optimization (TLBO) algorithm \cite{rao_teachinglearning-based_2012} and Latin hypercube sampling (LHS) are used to sample high- and low-fidelity infills, respectively. For both approaches, the surrogate is updated after each infill step and the procedure continues until it converges to an optimal solution in close agreement with the solver results. \par

\begin{table}[!ht]
	\centering
    \begin{tabular}{cc|cc|cc|cc}
    \hline\\[-1.0em]
    Ref. & Problem & \multicolumn{2}{c|}{DOEs} & \multicolumn{2}{c|}{Total infills} & Infill & Cost \\
    & dimension & LF & HF & LF & HF & strategy & ratio \\
    \hline
    \hline\\[-1.0em]
    \multicolumn{8}{c}{Multi-fidelity single-objective optimization with MFDNN}  \\
    \hline\\[-1.0em]
        Zhang et al. \cite{zhang_multi-fidelity_2021} & 30 & 120 & 60 & 120 & 60 & PSO$^\star$ \& ED & 7 \\
        Wu et al. \cite{wu_multi-fidelity_2024} & 12 & 160 & 20 & 1280 & 160 & TLBO$^\star$ \& LHS & - \\ \\[-1.0em]
    \hline\\[-1.0em]
    \multicolumn{8}{c}{Multi-fidelity multi-objective BO with AR1}  \\
    \hline\\[-1.0em]
        He et al. \cite{he_efficient_2024} & 144 & - & 720 & - & 150 & GEIM+ & 10 \\
        Mourousias t al. \cite{mourousias_novel_2024} & 30 & 2000 & 274 & \multicolumn{2}{c|}{9} & VFEIM+ & 30 \\
        Charayron et al. \cite{charayron_towards_2023} & 20 & 20 & 10 & \multicolumn{2}{c|}{10} & MPI+ & 116 \\
        Matar \cite{matar_camille_analysis_2024, matar2025cost} & 4 & 41 & 5 & 15 & 2 & PI \& LCB & 30 000 \\ \\[-1.0em]
    \hline
    \end{tabular}
    \caption{Main characteristics of multi-fidelity optimization methods with various adaptive infill strategies. The "Total infills" column indicates the total number of infill samples for each level of fidelity. In the infill strategy column, the "$\star$" symbol indicates that the high-fidelity sample is selected as the optimization algorithm best candidate. The "+" symbol indicates a variant of an existing strategy.}
    \label{tab:opt}
\end{table}

Although \cite{wu_multi-fidelity_2024} gives an ASO example where MFDNN outperformed single-fidelity kriging, both models were combined with a non-Bayesian infill strategy. It is therefore difficult to know whether this conclusion would hold if the kriging model was used in its multi-fidelity form and combined with a Bayesian infill strategy.\par 

Due to the encouraging results recently obtained with the MFDNN model \cite{zhang_multi-fidelity_2021, wu_multi-fidelity_2024} and the continuing popularity of the AR1 model despite its seniority, we selected these two methods for further comparison in the context of multi-fidelity ASO with adaptive infill under a limited number of high-fidelity samples.

\subsection{Shape parametrization and data reduction} 
In ASO, the shape under study is typically defined by a discretization of hundreds or thousands of surface points.
These, in turn, are parametrized by tens (in 2D) or hundreds (in 3D) of design variables that form the problem dimension. Shape parametrization methods are reviewed in \cite{skinner_state---art_2018}, which divides them into constructive and deformative models. Constructive models include various kinds of spline methods such as Bezier spline, basis spline and non-uniform rational basis spline. On the other hand, deformative models include free-form deformation (FFD) \cite{sederberg_free-form_1986}, class shape transformation \cite{kulfan_universal_2008} and Hicks-Henne bump function \cite{hicks_wing_2012}. As explained in \cite{skinner_state---art_2018}, the simplest methods for CFD-based optimization are deformative methods, and FFD is particularly suitable for high-fidelity optimization where the baseline geometry is already sufficiently close to the optimal solution. Since the optimization convergence speed and the minimum number of data required to train surrogate models are increasing functions of the problem dimensions, several data reduction methods have been proposed to reduce the parametrization size. Active subspace techniques use gradient evaluations to identify the most influential parameters of multivariate functions for which the principal directions of variability are not aligned with the coordinates of the input space \cite{constantine_active_2014}. Since then, both linear (e.g. POD) and non-linear (e.g. autoencoders, generative adversarial networks) gradient-free reduction techniques applied to the shape parametrization have become common practice \cite{li_machine_2022} and remain an active field of research \cite{cinquegrana_efficient_2017, wu_benchmark_2019, ullah_exploring_2020, hou_dimensionality_2022}. In \cite{ullah_exploring_2020, hou_dimensionality_2022}, principal component analysis (also sometimes referred to as proper orthogonal decomposition, POD, or singular value decomposition, SVD) is shown to simultaneously reduce the number of necessary samples to train surrogate models and to improve their modeling accuracy. A similar approach has been recently considered in \cite{matar_camille_analysis_2024, matar2025cost}. For all these reasons, an approach combining FFD and POD is adopted herein.

\subsection{Discretization error control}
When dealing with CFD-based optimization, the challenge of mesh management is of first importance since function objective evaluation can only be as good as the discretization used to compute it. Poor choices in mesh resolution and numerical solvers can lead to errors that propagate through the surrogate model \cite{van2018goal}, ultimately resulting in a misguided optimal design. Indeed, Cinnella and Congedo \cite{cinnella_convergence_2013} showed for a transonic airfoil single optimization problem that the numerical errors in the representation of the objective function makes the genetic algorithm convergence harder. 
While optimizations on fine grids or the use of Richardson extrapolation can mitigate the problems, none of them guarantees optimal mesh quality for all the designs explored during the optimization.\par

A more rigorous way of controlling numerical errors and ensuring mesh quality is to use automatic mesh adaptation techniques.
Mesh adaptation methods are of two kinds: feature-based mesh adaptation, which aims to provide the best mesh to compute the characteristics of a given sensor (e.g. Mach field), and the adjoint-based (or goal-oriented) mesh adaptation, which targets the best mesh to observe a given scalar quantity of interest (e.g. drag coefficient) \cite{alauzet_decade_2016, belme2019priori,alauzet_feature-based_2021}. Despite their successful application to both external \cite{alauzet_feature-based_2021} and internal \cite{alauzet_periodic_2022} flow computations, mesh adaptation tools are still not widely available in CFD codes, and their use in ASO is quite rare.
For instance, John et al. \cite{john_using_2020} showed the benefit of feature-based mesh adaptation for the aerodynamic design of turbomachinery configurations with fully structured block meshes and single objective adjoint-based optimization. Chen et al. investigated single \cite{chen_discretization_2019} and multi point \cite{chen_variable-fidelity_2020} ASO with adjoint-based mesh adaptation. In both papers, particular attention is paid to avoiding over-refinement of undesirable designs according to the estimated error in the objective function, thus avoiding unnecessary computations. To reduce the computational costs, an attractive alternative consists in combining mesh adaptation with multi-fidelity optimization. This was recently done by Wackers et al. \cite{wackers_multi-fidelity_2020, wackers_improving_2022}, who showed the effectiveness of combining multi-fidelity radial basis functions with mesh adaptation and adaptive infills for the single-objective optimization of hydrodynamic shapes parametrized with only one or two design variables. However, to the authors' knowledge, such a strategy has not been applied to multi-objective ASO in high-dimensional spaces. \\ \par

\section{Methodology}\label{sec:methodology}
In the following we consider two multi-fidelity  methodologies for solving multi-objective optimization problems. In particular, we focus on two-objective problems. Given a design space $\Omega\subset\mathbb{R}^d$, with $d$ the number of design variables, and $n_\textrm{obj}$ objective functions $J_k$ with $k \in \llbracket1,n_\textrm{obj}\rrbracket$, our goal is to find a set $PS$ of Pareto-optimal individuals, i.e. designs $x^*$ satisfying the Pareto dominance conditions \cite{gaspar-cunha_modified_2015}:
\begin{equation}
    \forall x \in \Omega\, \left\{
    \begin{array}{l}
        \forall k \in \llbracket 1, n_\textrm{obj}\rrbracket,\, J_k(x^*)\le J_k(x), \\[9pt]
         \exists j \in \llbracket 1, n_\textrm{obj}\rrbracket,\, J_j(x^*) < J_j(x).
    \end{array}\right. 
\end{equation}
In order to speed-up the optimization, in the following we seek for surrogate models that approximate the input/output relationships $J_k: x \mapsto J_k(x)$ using information from a set of low-fidelity and high-fidelity evaluations of the objectives. Specifically, our goal is to get the best possible approximation of $PS$ using the least  number of (presumably very) expensive high-fidelity function calls.

\subsection{Multi-fidelity surrogate-based optimization}\label{sec:mfo}
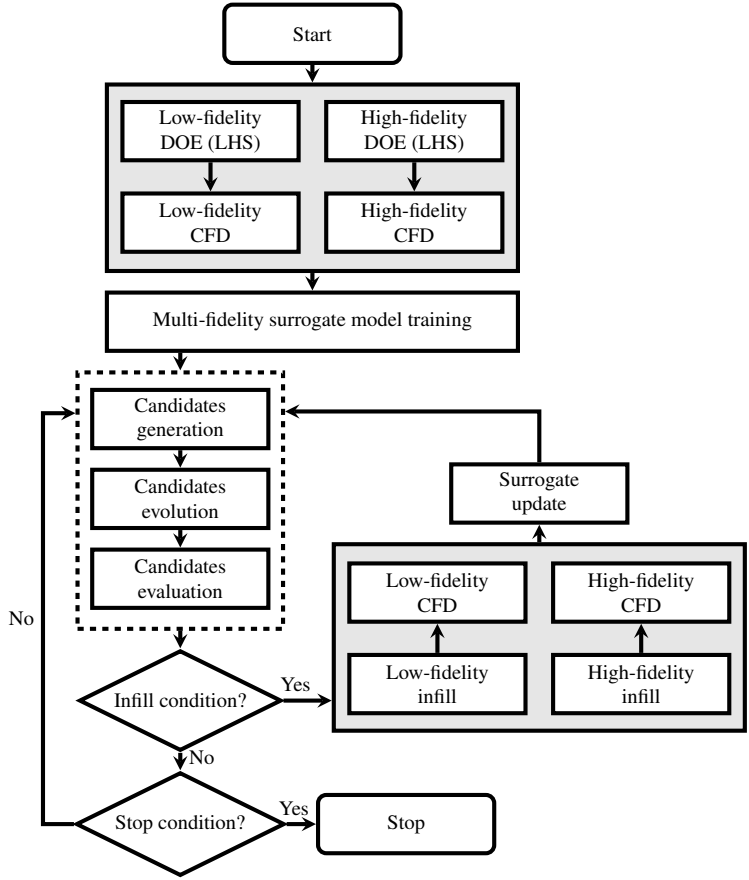
\begin{figure}[!htb]
    \centering
    \resizebox{0.6\textwidth}{!}{\begin{tikzpicture}[node distance=2cm]
\node (start) [rectangle-b] {Start};
\node (bbox) [rectangle-v, below=3 mm of start] {};
\node (lf-lhs) [rectangle-o, align=center, below right=3mm and 1.8cm of bbox.north west, anchor=north] {Low-fidelity\\DOE (LHS)};
\node (hf-lhs) [rectangle-o, align=center, below left=3mm and 1.8cm of bbox.north east, anchor=north] {High-fidelity\\DOE (LHS)};
\node (lf-cfd) [rectangle-g, align=center, below=5mm of lf-lhs] {Low-fidelity\\CFD};
\node (hf-cfd) [rectangle-g, align=center, below=5mm of hf-lhs] {High-fidelity\\CFD};
\node (sm-box) [large-rectangle-o, below=3mm of bbox] {Multi-fidelity surrogate model training};
\node (algo) [rectangle-h, below right=3mm and -0.5cm of sm-box.south west]{};
\node (pop-gen) [rectangle-o, align=center, below=3mm of algo.north] {Candidates\\generation};
\node (pop-evo) [rectangle-o, align=center, below=3mm of pop-gen] {Candidates\\evolution};
\node (pop-eva) [rectangle-o, align=center, below=3mm of pop-evo] {Candidates\\evaluation};
\node (in-con) [diamond-g, align=center, below=3mm of algo] {Infill condition?};
\node (bbox-in) [rectangle-v, above right=-1cm and 1.7cm of in-con] {};
\node (in-lf) [rectangle-o, align=center, above right=3mm and 1.8cm of bbox-in.south west, anchor=south] {Low-fidelity\\infill};
\node (in-hf) [rectangle-o, align=center, above left=3mm and 1.8cm of bbox-in.south east, anchor=south] {High-fidelity\\infill};
\node (in-lf-cfd) [rectangle-g, align=center, above=5mm of in-lf] {Low-fidelity\\CFD};
\node (in-hf-cfd) [rectangle-g, align=center, above=5mm of in-hf] {High-fidelity\\CFD};
\node (st-con) [diamond-g, align=center, below=3mm of in-con] {Stop condition?};
\node (sm-upd) [rectangle-o, align=center, above=3mm of bbox-in] {Surrogate\\update};
\node (stop) [rectangle-b, right=5mm of st-con] {Stop};
\draw [arrow] (start) -- (bbox);
\draw [arrow] (lf-lhs) -- (lf-cfd);
\draw [arrow] (hf-lhs) -- (hf-cfd);
\draw [arrow] (bbox) -- (sm-box);
\draw [-arrow] (algo) -- (algo |- sm-box.south);
\draw [arrow] (pop-gen) -- (pop-evo);
\draw [arrow] (pop-evo) -- (pop-eva);
\draw [arrow] (algo) -- (in-con);
\draw [arrow] (in-con) -- (st-con) node [right, pos=0.25] {No};
\draw [arrow] (in-con.east) -- (in-con.east -| bbox-in.west) node [above, pos=0.25] {Yes};
\draw [arrow] (in-lf) -- (in-lf-cfd);
\draw [arrow] (in-hf) -- (in-hf-cfd);
\draw [arrow] (bbox-in) -- (sm-upd);
\draw [arrow] (sm-upd) |- ([yshift=5.25cm]algo);
\draw [arrow] (st-con) -- (stop) node [above,pos=0.25] {Yes};
\draw [arrow] (st-con.west) -- ++(-15pt,0pt) |- ([yshift=1.5cm]algo.west) node [left, pos=0.25] {No};
\end{tikzpicture}}
    \caption{Workflow of the multi-fidelity surrogate assisted optimization framework. The dashed box represents the optimization algorithm that may be skipped in case of Bayesian optimization. The gray boxes represent multi-step processes including candidate sampling, geometry deformation, mesh generation and low- or high-fidelity CFD simulations.}
    \label{fig:aso}
\end{figure}

In the present Section we introduce two multi-objective surrogate-based optimization strategies with adaptive infills. The first one relies on a AR1 multi-fidelity surrogate coupled with Bayesian optimization whereas the second one uses a multi-fidelity neural network combined with a genetic algorithm. The high level structure of the framework\footnote{The \textit{github} repository will be made available upon publication.} is illustrated in \autoref{fig:aso} and each component (optimization algorithm, surrogate assistance, CFD fidelity, adaptive infill, data reduction) is configurable such that different optimization approaches can be compared.

\subsubsection{Bayesian multi-fidelity optimization }
The first approach under consideration combines a multi-fidelity Gaussian process, and a Bayesian optimization algorithm with adaptive infills.

In its bi-fidelity form, the co-kriging model (AR1) which relies on Kennedy and O'Hagan linear autoregressive fusion scheme \cite{kennedy_predicting_2000} relates a high-fidelity model $f_{hf}(\cdot)$ to a low-fidelity model $f_{lf}(\cdot)$ through the expression:
\begin{equation}
    f_{hf}(x) = \rho\, f_{lf}(x) + \delta(x).
\end{equation}
The scaling factor $\rho$ may depend on $x$ but is usually assumed to be constant, while the additive bias $\delta(\cdot)$ is a Gaussian process that models the difference between the two fidelities. In this work, we use the AR1 implementation of the \textit{smt} package\footnote{\textit{smt} is an open source Python library hosted on github (see \url{https://smt.readthedocs.io/en/latest/}).}  \cite{saves_smt_2024}, which follows the recursive formulation of \cite{le_gratiet_recursive_2014}. Such algorithm relies on a nested design of experiments (DOE) for the two levels of fidelity, i.e. high- and low-fidelity samples must be evaluated at the same locations in the parameter space. The nested structure allows to express the surrogate variance in a closed form. A benefit of this approach is that when dealing with $L$ levels of fidelity, the predictive mean $\mu_l(\cdot)$ and variance $\sigma_l^2(\cdot)$ of each fidelity level $l \in \llbracket 2; L \rrbracket$ can be expressed recursively as follows \cite{charayron_towards_2023}:
\begin{align*}
    & \mu_l(x) = \rho_{l-1}\, \mu_{l-1}(x) + \mu_{\delta_l}(x), \\
    & \sigma_l^2 = \rho_{l-1}^2\, \sigma_{l-1}^2 + \sigma_{\delta_l}^2,
\end{align*}
(with $\mu_{\delta_l}$ and $\sigma_{\delta_l}^2$ the mean and variance of the bias term, respectively), so that each fidelity level can be trained separately. \\ \par

With AR1 models, a separate surrogate model is trained for each objective function. Then, Bayesian optimization requires the definition of a suitable acquisition function that updates the best-performing designs using information from the surrogates. Hereafter, an optimization strategy is devised by mixing ideas from the last two approaches in \autoref{tab:opt} and described in \autoref{alg:binfill}. \par

\begin{algorithm}[H]
\caption{Bayesian adaptive infill optimization.}\label{alg:binfill}
\begin{algorithmic}
    \State Compute low- and high-fidelity initial DOEs
    \State Train AR1 models $\mu_1$ and $\mu_2$
    \For {$i=1$ to $n_{iter}$}
        \State $x_{hf}=\underset{x}{\text{argmax}} \Bigl\{\alpha_{WB2S,MPI}(x) \Bigr\}$
        \State $x_{lf}=x_{hf} \cup \underset{x}{\text{argmin}} \Bigl\{LCB_1(x) \Bigr\} \cup \underset{x}{\text{argmin}} \Bigl\{LCB_2(x) \Bigr\}$
        \For {$j=1$ to $n_{lf}$}
            \State $X_{lf}=X_{lf} \cup x_{lf}$
            \State $x_{lf}=x_{lf} \cup \underset{x}{\text{argmax}} \Bigl\{\min\, d(x, X_{lf}) \Bigr\}$
        \EndFor
        \State Compute $y_{hf}$ and $y_{lf}$
        \State Re-train AR1 models $\mu_1$ and $\mu_2$
    \EndFor
    \State Run NSGA-II with $\mu_1$ and $\mu_2$
\end{algorithmic}
\end{algorithm}
First, the low- and high-fidelity initial datasets are generated and AR1 models are trained. Then, at each iteration of the optimization, a high-fidelity infill sample is selected based on a regularized infill criterion. As explained in \cite{grapin_constrained_2022}, regularization techniques helps to circumvent the ill-posedness of infill problems. Specifically, given any multi-objective acquisition function $\alpha_f(\cdot)$, the next infill sample is obtained by maximizing:
\begin{equation}
	\alpha_{WB2S,f}(x) = \gamma\, \alpha_f(x) - \psi(\mu_f(x)).
\end{equation}
where $\alpha_{WB2S,f}$ is a regularized acquisition function, $\gamma$ is a real parameter, $\mu_f(\cdot)$ is the collection of the Gaussian processes for each objective and $\psi(\cdot)$ is a so-called scalarization operator (e.g. max or sum). Similar to what was done in \cite{charayron_towards_2023}, we chose $\psi$ to be the sum function, while the initial acquisition function $\alpha_f$ is taken as the minimal probability of improvement (MPI) \cite{rahat_alternative_2017}, defined as:
\begin{equation}\label{eq:pi}
	MPI(x) = \underset{u \in PS}{\min} \left(1 - \prod_{k=1}^{n_{obj}}\Phi\left(\frac{\mu_k(x) - \hat{y}_k(u)}{\sigma_k(x)} \right) \right).
\end{equation}
In the preceding expression, $\Phi(\cdot)$ is the normal cumulative distribution function while $\mu_{k}(\cdot)$ and $\sigma_{k}(\cdot)$ denote the predictive mean and standard deviation of objective $k$. The true high-fidelity response for objective $k$ is denoted $\hat{y}_k(\cdot)$ and $u\in PS$ form the current high-fidelity Pareto set.

In order to satisfy the nested dataset property, the infill point selected by the acquisition function is computed using both the high-fidelity and the low-fidelity model. In addition, two additional low-fidelity infill samples are selected by minimizing the lower confidence bound (LCB) of each objective as suggested in \cite{matar_camille_analysis_2024, matar2025cost}:
\begin{equation}\label{eq:lcb}
	LCB_k(x) = \mu_k(x) - \sigma_k(x),\, k \in \llbracket1,n_\text{obj}\rrbracket.
\end{equation}
To favor diversity, additional low-fidelity candidates are selected by maximizing the minimal Euclidean distance from \cite{zhang_multi-fidelity_2021}:
\begin{equation}\label{eq:ed}
	\max \min d(x, X_{lf}), 
\end{equation}
where $d(\cdot, \cdot)$ denotes the Euclidean distance between the new point $x$ and the current low-fidelity DOE $X_{lf}$. 
At each infill step, all three optimization sub-problems resulting from \autoref{eq:pi}, \ref{eq:lcb} and \ref{eq:ed} are solved with particle swarm optimization. At that point, the surrogate models of each objective are updated and the procedure is repeated until the specified number of iterations based on the computational budget is reached. Lastly, an extra post-processing step is performed by running the NSGA-II genetic algorithm \cite{nsgaII} with the final version of the surrogate models. This has for objective to provide a denser estimate of the Pareto front \cite{charayron_towards_2023}. In the following, it will be referenced to as the predicted Pareto.

\subsubsection{Non-Bayesian multi-fidelity optimization}
The second considered strategy combines a multi-fidelity deep neural network (MFDNN) with a multi-objective genetic algorithm.

\begin{figure}[!htb]
    \centering
        \begin{tikzpicture}
            \draw (0, 0) node[inner sep=0] {\includegraphics[trim=2.5cm 2.5cm 2.5cm 2.5cm,clip,width=0.8\textwidth]{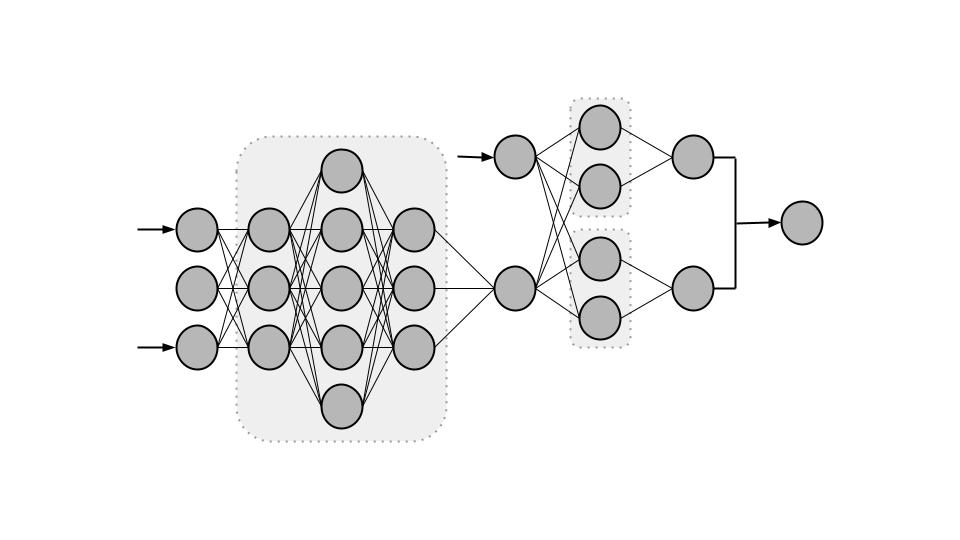}};
		    \draw (-5.5, 0.9) node {$x_{L,1}$};
            \draw (-5.5, -0.1) node {$\vdots$};
            \draw (-5.5, -1) node {$x_{L,d}$};
            \draw (-2.2, 2.4) node {$NN_L$};
            \draw (0.6, -0.3) node {$y_{L}$};
            \draw (-0.2, 3) node {$x_{H,1}$};
            \draw (-0.2, 2.6) node {$\vdots$};
            \draw (-0.2, 2.1) node {$x_{H,d}$};
            \draw (2, -1.6) node {$NN_{H_2}$};
            \draw (2, 3.05) node {$NN_{H_1}$};
            \draw (4.4, 2.1) node {$\beta \mathcal{F}_l$};
            \draw (4.5, -0.7) node {$(1 - \beta) \mathcal{F}_{nl}$};
            \draw (5.2, 0.75) node {$y_{H}$};
        \end{tikzpicture}
    \caption{MFDNN architecture reproduced from \cite{wu_multi-fidelity_2024}.}
    \label{fig:mfdnn}
\end{figure}
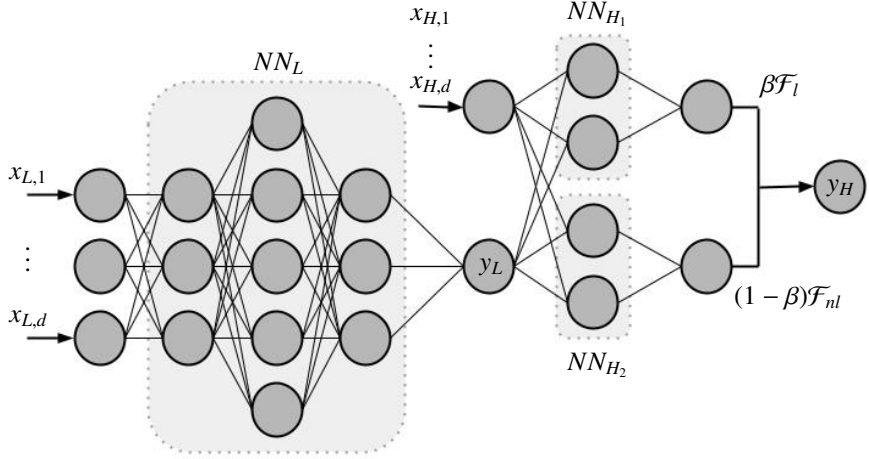

The MFDNN model was initially introduced by Meng et al. \cite{meng_composite_2020}. In contrast with the linear correlation used in AR1 models, it relies on the combination of a linear neural network $\mathcal{F}_l(\cdot)$ and a non-linear neural network $\mathcal{F}_{nl}(\cdot)$ according to the following equation:
\begin{equation}\label{eq:mfddn}
    f_{hf}(x) = \beta \mathcal{F}_l(x, y_L) + (1 - \beta) \mathcal{F}_{nl}(x, y_L),
\end{equation}
where the correlation coefficient $\beta \in [0; 1]$ can be learned \cite{zhang_multi-fidelity_2021} or fixed \cite{wu_multi-fidelity_2024}. The network architecture is illustrated in \autoref{fig:mfdnn}, where the low-fidelity model is approximated with a neural network $NN_L(x_L, \theta_0)$,  combined with two other networks $NN_{H_1}(x, y_L, \theta_1)$, $NN_{H_2}(x, y_L, \theta_2)$ that are used to approximate the correlation between the low- and high-fidelity data.
The parameters $\theta_i$ for each network are trained by minimizing the loss function:
\begin{align}
    & \mathcal{L}oss = \mathcal{L}oss_{y_L} + \mathcal{L}oss_{y_H} + \lambda \sum \theta_i^2, \\
    &\notag \mathcal{L}oss_{y_L} = \frac{1}{N_{y_L}} \sum_{i=1}^{N_{y_L}} \left(\vert \hat{y}_L - y_L \vert^2\right) \\
    &\notag \mathcal{L}oss_{y_H} = \frac{1}{N_{y_H}} \sum_{i=1}^{N_{y_H}} \left(\vert \hat{y}_H - y_H \vert^2 \right),
\end{align}
where $\hat{y}_{L}$ and $\hat{y}_{H}$ are the true low- and high-fidelity values, while $y_{L}$ and $y_H$ are the predicted low- and high-fidelity values. While not required by the formulation of MFDNN models, the use of nested DOEs to construct the training sets was suggested in \cite{wu_multi-fidelity_2024}. The effect of this choice is therefore investigated in the next section. The MFDNN  was implemented using the \textit{Pytorch}\footnote{\textit{Pytorch} is an open source Python library hosted on github (see \url{https://pytorch.org/}).} library. \par

Since the MFDNN surrogate is deterministic, a non-Bayesian optimization strategy is selected in conjunction with this model, described in \autoref{alg:nbinfill}. Unlike AR1 models, MFDNNs can easily be generalized to any number of objectives by extending the last layer of each of its sub-networks. As a drawback, no Bayesian consideration can be used to guide the selection of infill samples and update the optimization. For this reason, the MFDNN model is coupled with the NSGA-II genetic algorithm to obtain intermediary estimates of the Pareto front with increasing reliability. At each iteration, the best-performing candidates are extracted from the Pareto set. The candidate offering the best compromise is recomputed with both the low- and high-fidelity solvers, hence ensuring the nested property of the datasets. The best candidates with respect to each objective are recomputed with the low-fidelity solver only. Similarly to the Bayesian approach, additional low-fidelity samples can be selected by maximizing the minimal Euclidean distance. Once the specified number of iterations is reached, an extra run of NSGA-II is performed to densify the Pareto front. \\ \par

\begin{algorithm}[H]
\caption{Non-Bayesian adaptive infill optimization.}\label{alg:nbinfill}
\begin{algorithmic}
    \State Compute low- and high-fidelity initial DOEs
    \State Train the MFDNN model
    \For {$i=1$ to $n_{iter}$}
        \State Run NSGA-II with the MFDNN model
        \State $x_{hf}$ taken as the estimated Pareto center
        \State $x_{lf}$ taken as the estimated Pareto ends and center
        \For {$j=1$ to $n_{lf}$}
            \State $X_{lf}=X_{lf} \cup x_{lf}$
            \State $x_{lf}=x_{lf} \cup \underset{x}{\text{argmax}} \Bigl\{\min\, d(x, X_{lf}) \Bigr\}$
        \EndFor
        \State Compute $y_{hf}$ and $y_{lf}$
        \State Re-train the MFDNN model
    \EndFor
    \State Run NSGA-II with the MFDNN model
\end{algorithmic}
\end{algorithm}

\subsection{Pareto front evaluation metrics}\label{sec:metrics}
In the Results section (see \autoref{sec:cascade}), the performances of  different optimization strategies are compared in the light of three multi-objective metrics. 

The inverted generational distance (IGD) is defined as \cite{gaspar-cunha_modified_2015}:
\begin{equation}
	IGD(A) = \frac{1}{\vert Z \vert} \left(\sum_{j=1}^{\vert Z \vert} \hat{d}_j^p \right)^{1/p},
\end{equation}
where $Z$ is the reference Pareto front, $A$ is an experiment resulting Pareto set and  $\hat{d}_j$ is the Euclidean distance from $z_j \in Z$ to its nearest objective vector A. The inverted generational distance plus (IGD+) is defined as \cite{gaspar-cunha_modified_2015}:
\begin{equation}
	IGD^{+}(A) = \frac{1}{\vert Z \vert} \left( \sum_{j=1}^{\vert Z \vert} d_j^{+}{}^2 \right)^{1/2},
\end{equation}
where $d_i^{+}=\max(a_i - z_i, 0)$ is the modified distance from $a_i \in A$ to its nearest reference point $z_i \in Z$. A major difference between these two metrics is that $IGD(\cdot)$ is not Pareto compliant whereas $IGD^{+}(\cdot)$ is weakly Pareto compliant \cite{falcon-cardona_construction_2022}. Weak Pareto compliance means that if $A_1$ dominates $A_2$, then $IGD^{+}(A_1) < IGD^{+}(A_2)$ \cite{falcon-cardona_construction_2022}. The last indicator is the hypervolume (HV), a quantity that measures the volume of objective space dominated by the Pareto front, bounded by a reference point (here, the baseline configuration).

\subsection{Geometry parametrization and problem dimension reduction}
The geometry parametrization is a key component of the optimization strategy, since it determines the number of design variables and hence the dimension of the search space. Here, we target the optimization of two-dimensional airfoil profiles for turbomachinery configurations, which can require a significant number of design variables in order to control various parameters of engineering interest. A number of parametrization techniques have been considered in the literature, aiming at striking a compromise between flexibility and parsimony of the representation (a recent contribution can be found in \cite{voss2025differentiated}).
In the aerodynamic optimization use case considered later in this paper, we follow previous work by \cite{bufi2017robust,serafino2023multi}, who used Free-Form deformation (FFD) to efficiently describe a wide variety of turbine blade sections using a small number of control parameters.\\ \par

\begin{figure}[!htb]
\centering
  \includegraphics[width=\linewidth]{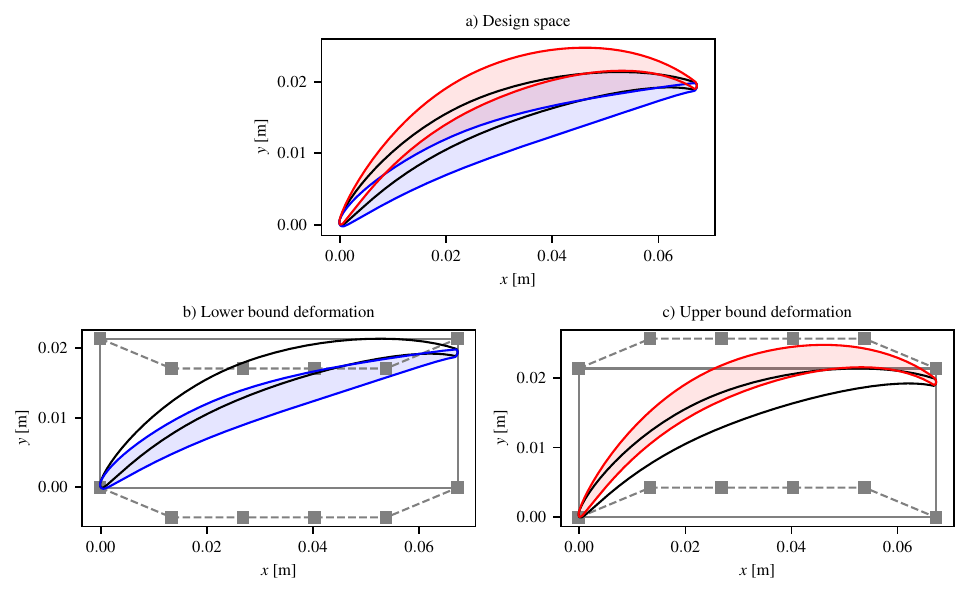}
  \caption{LRN cascade design space (a) and extreme deformations (b-c) for 8 FFD control points.}
  \label{fig:lrn-cascade-ffd}
\end{figure}

First introduced by Sederberg in 1986 \cite{sederberg_free-form_1986}, FFD consists in embedding a solid geometry inside a hull volume parametrized by control points. These points are then perturbed to deform the volume which results in geometric changes of the solid geometry. A 2D FFD of the DLR low Reynolds number (LRN) compressor cascade optimized in \autoref{sec:cascade} is illustrated in \autoref{fig:lrn-cascade-ffd}, using 8 control points. \par

\begin{figure}[!htb]
    \centering
    \includegraphics[width=\linewidth]{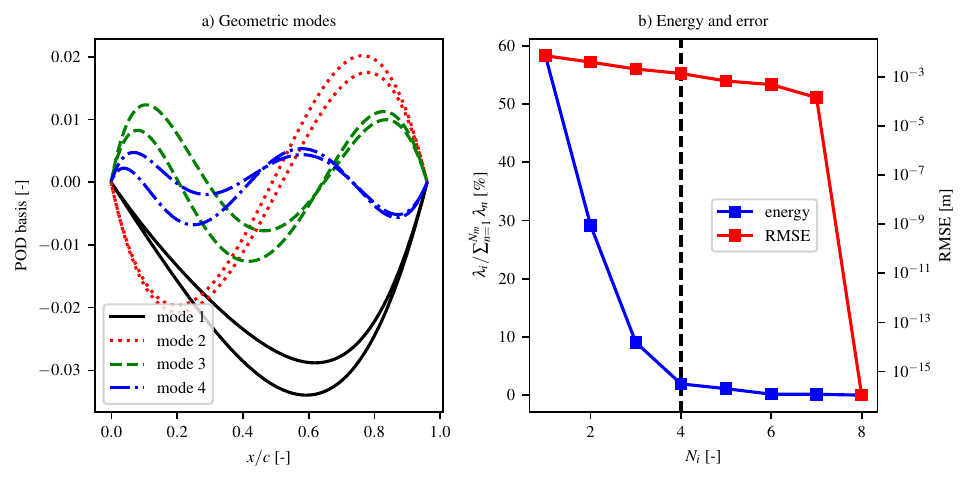}
    \caption{LRN cascade POD 4 principal geometric modes (a), energy and reconstruction error as a function of the number of modes (b). The black dashed line indicates the selected mode number $N_m=4$ which conserves 98.5\% of the modal energy.}
    \label{fig:lrn-cascade-pod}
\end{figure}

To reduce even further the number of design points, a Proper Orthogonal Decomposition (POD) of the geometrical space is also considered. The details of this reduction technique are given in \ref{app:pod}. The POD decomposition maps the initial $d$-dimensional  space defined by the FFD control parameters onto a reduced latent space of dimension $d^\star < d$. For the LRN cascade geometry at stake, the geometric modes, the energy and the reconstruction error are represented in \autoref{fig:lrn-cascade-pod}. 
It is worth noting that although FFD control point displacements have a physical meaning, the modal coefficients in the projected space do not. An immediate consequence is that the reduced design variables are harder to interpret. Another consequence is that the extreme profiles are not conserved in the reduced space (see \ref{app:pod}). \par

\subsection{CFD solver and feature-based anisotropic mesh adaptation}
Throughout this work, both low- and high-fidelity solutions are computed with the CFD solver \textit{wolf}, a vertex-centered mixed finite-volume / finite-element Navier-Stokes solver for both 2D or 3D unstructured meshes. \textit{Wolf} is designed for anisotropic metric-based mesh adaptation. It can then be coupled with \textit{feflo.a}, a generic purpose adaptive metric-based mesh generator, to generate adapted 2D or 3D meshes. A synthetic description of \textit{wolf} and \textit{feflo.a} is available in \cite{alauzet_new_2024} and a detailed description is provided in \cite{alauzet_periodic_2022}. Hereafter, we briefly recall the mesh adaptation strategy. \par

Starting off from a mesh $\mathcal{H}$, the goal of mesh adaptation is to compute an optimal mesh $\mathcal{H}_{opt}$ by minimizing a given error estimate $E$, for a fixed number of elements $\mathcal{N}_e$:
\begin{equation}\label{eq:adap}
	\mathcal{H}_{opt} = \underset{\mathcal{N}_e}{\mathrm{argmin}}\, E(\mathcal{H}).
\end{equation}
The metric-based mesh adaptation approach recasts the discrete mesh $\mathcal{H}$ into a continuous Riemannian metric space where the continuous reformulation of \autoref{eq:adap} can be solved analytically for a fixed continuous mesh complexity $\mathcal{N}_c \propto \mathcal{N}_e$. In the specific case of feature-based mesh adaptation, the goal is to minimize the interpolation error  of a given ``sensor field''. This problem can be solved analytically in the metric space as a function of the Hessian of the sensor field  \cite{alauzet_new_2024}. In the present computations we used the Mach number as a sensor because it guarantees the convergence of the solution and automatically captures complex flow features such as boundary layers, wakes and shock waves \cite{alauzet_periodic_2022}. Details of the mesh adaptation algorithm can be found in the same reference. \par

The \textit{wolf} solver uses the Spalart-Allmaras (SA) RANS model \cite{allmaras_modi_2012} to evaluate the turbulent stresses. However, RANS models are not well suited to capture laminar/turbulent transition phenomena that often occur in turbomachinery configurations. For this reason we consider the SA-BCM version, which takes into account the intermittency function $\gamma$, a local indicator of whether the flow should be considered as turbulent or not \cite{cakmakcioglu_revised_2020, mura_revised_2020}. As opposed to more sophisticated transition models, this formulation is algebraic and does not require the resolution of additional transport equation, while still providing reasonably accurate results, as shown in \autoref{sec:cascade}.

\section{Preliminary tests}\label{sec:validation}

\subsection{Multi-fidelity surrogates evaluation: 1D test functions}
\begin{figure}[!htbp]
\centering
\includegraphics[width=\textwidth]{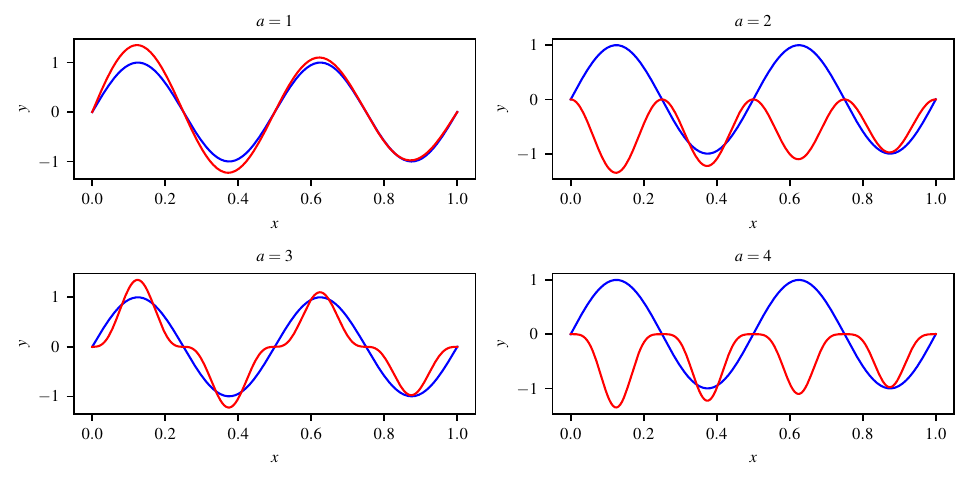}
\caption{1D analytical test functions from \cite{brevault_overview_2020}. The red and blue curves respectively correspond to the high- and low-fidelity functions.}
\label{fig:brevault1d}
\end{figure}

Gaussian process based multi-fidelity models were compared by Brevault et al. in \cite{brevault_overview_2020} with the aim of evaluating their strengths and weaknesses depending on the characteristics of several analytical and aerospace problems. Their study includes both simple correlation models such as AR1, and more advanced models such as the non-linear auto-regressive Gaussian process and deep Gaussian process. Their results suggest that when the number of available high-fidelity samples is limited, simple models tend to perform better than more complex approaches. Therefore, to get a first idea of how MFDNN and AR1 models compare, a preliminary comparison of these surrogates is made using the 1D test functions used in \cite{brevault_overview_2020}:
\begin{align*}
    & f_{lf}(x) = \sin(4 \pi x), \\
    & f_{hf}(x) = \left(\frac{x}{2} - \sqrt{2}\right) \sin(4 \pi x + a \pi)^a, \\
    & \notag x\in[0; 1] \text{ and } a\in \llbracket 1; 4\rrbracket.
\end{align*}
The subscripts $hf$ and $lf$ denote the high-fidelity function and its low-fidelity approximation, respectively. The parameter $a$ controls the correlation between the two fidelities by introducing phase and amplitude differences, as shown in \autoref{fig:brevault1d}. \par

For each value of $a$, DOEs of respectively 10 and 40 high- and low-fidelity samples are drawn through Latin hypercube sampling (LHS) and are used for training the surrogate. The accuracy of the models is then computed by means of the root-mean squared error (RMSE) between the models and the high-fidelity solution evaluated at 1000 random points. For each configuration, the measures are repeated ten times using different training and evaluation DOEs, and the mean, as well as the minimum and maximum errors across the repetitions are reported. \par

In the present series of numerical experiments, the AR1 scaling factor is learned. The considered MFDNN architectures are those of Zhang et al. \cite{zhang_multi-fidelity_2021} and Wu et al. \cite{wu_multi-fidelity_2024} as given in \autoref{tab:mfdnn}. For both configurations $\beta$ is learned as well, the learning rate is set to $10^{-4}$, the loss function is reduced with Adam optimizer \cite{kingma2014adam} and the network parameters are initialized with Xavier's initialization method \cite{pmlr-v9-glorot10a}. This means that for each repetition, the networks start from different initial states. \par

\begin{figure}[!htb]
\centering
\begin{minipage}[t]{.49\textwidth}
  \centering
  \includegraphics[width=\linewidth]{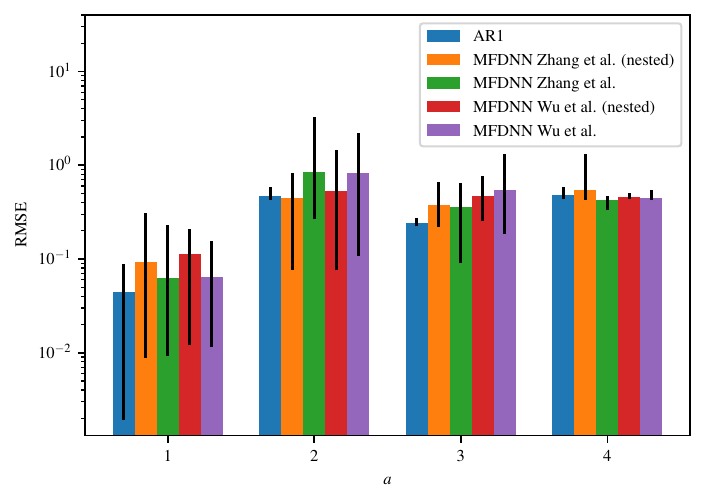}
  \caption{RMSE of the multi-fidelity models for the 1D test functions with different values of the parameter $a$. Each model is trained ten times using different DOEs of respectively 40 and 10 low- and high-fidelity samples. The error bars denote the min/max RMSE registered across the repetitions.}
  \label{fig:mfdnn-cokg-a}
\end{minipage}\hfill
\begin{minipage}[t]{.49\textwidth}
  \includegraphics[width=\linewidth]{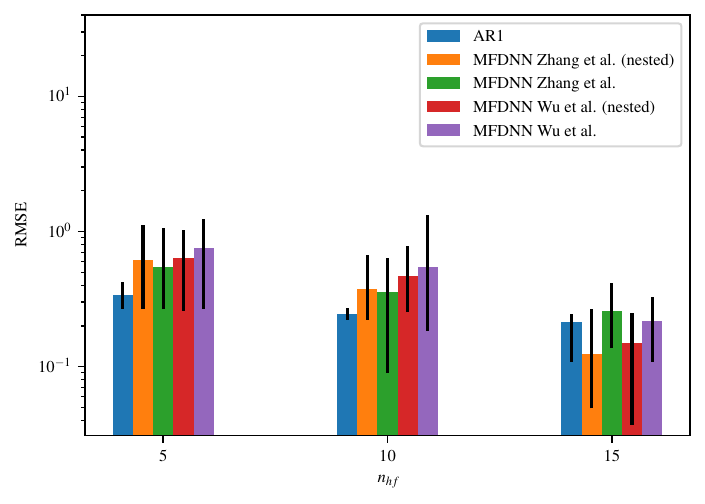}
  \caption{RMSE of the multi-fidelity models for the 1D test function defined by $a=3$ and different numbers of high-fidelity samples $n_{hf}$. Each model is trained ten times using different DOEs and the error bars denote the min/max RMSE registered across the repetitions.}
  \label{fig:mfdnn-cokg-hf}
\end{minipage}
\end{figure}

The histogram in \autoref{fig:mfdnn-cokg-a} reports the mean errors obtained for the AR1 and various versions of the MFDNN, with error bars corresponding to the min/max RMSE values. For this simple 1D problem, all models perform reasonably well for any degree of correlation (related to the chosen value of $a$) between the low- and high-fidelity models, even if smaller errors are registered for the well-correlated case $a=1$. Of note, the MFDNN variations around the mean RMSE are much higher than those found for the AR1, so that for all $a>1$, there is at least one configuration for which the MFDNN outperforms AR1. However, the two models do not differ much on average. \par

Considering the non-linearly correlated functions given by $a=3$, \autoref{fig:mfdnn-cokg-hf} shows that the MFDNN model tend to perform better than the AR1 when the number of high-fidelity point increases, but not in the low high-fidelity sample regime targeted in our study. The results also suggest that using nested DOEs improves the accuracy performances of the MFDNN. \par

\begin{table}[!ht]
    \centering
    \begin{tabular}{ccccc}
    \hline\\[-1.0em]
    Neural network & Layers & Neurons per layer & Activation & Weight decay \\
    \hline
    \hline\\[-1.0em]
    \multicolumn{5}{c}{MFDNN Zhang et al. \cite{zhang_multi-fidelity_2021}}  \\
    \hline\\[-1.0em]
        $NN_L$ & 6 & 32 & Tanh & $10^{-4}$ \\
        $NN_{H1}$ & 1 & 16 & - & - \\
        $NN_{H2}$ & 2 & 16 & Tanh & $10^{-4}$ \\
    \hline\\[-1.0em]
    \multicolumn{5}{c}{MFDNN Wu et al. \cite{wu_multi-fidelity_2024}}  \\
    \hline\\[-1.0em]
        $NN_L$ & 6 & 20 & Tanh & $10^{-5}$\\
        $NN_{H1}$ & 3 & 16 & - & $10^{-5}$ \\
        $NN_{H2}$ & 3 & 16 & Tanh & $10^{-5}$ \\
    \hline
    \end{tabular}
    \caption{Parameters of the MFDNN architectures.}
    \label{tab:mfdnn}
\end{table}

\subsection{Evaluation of surrogate-based optimization using Bayesian and non-Bayesian adaptive infills}
In this section, we carry out a preliminary assessment of the proposed optimization strategies for two test problems extracted from the popular so-called ZDT test suite initially proposed in \cite{zitzler2000comparison} for bi-objective optimization algorithms. Specifically, we focus on the ZDT1 and ZDT2 cases adapted to multi-fidelity optimization. 
\par
Given two objective functions $J_1$ and $J_2$, the ZDT1 problem is defined with the following low and high-fidelity functions \cite{charayron_towards_2023}:
\begin{equation*}
	\begin{array}{c}
	(J_1) \quad \left\{\begin{array}{l}
		f_{hf}(x) = x_1 \\[9pt]
		f_{lf}(x) = 0.9 x_1 + 0.1
	\end{array}\right. \qquad
	(J_2) \quad \left\{\begin{array}{l}
		f_{hf}(x) = u(x) v(x) \\[9pt]
		f_{lf}(x) = (0.8 u(x) - 0.2)(1.2 v(x) + 0.2) \\[9pt]
		u(x) = 1 + \frac{9}{d-1}\sum_{i=2}^{d} x_i \\[9pt]
		v(x) = 1 - \sqrt{\frac{x_1}{u(x)}}
	\end{array}\right.
	\end{array}
\end{equation*}
where $d$ is the problem dimension and $x\in \mathbb{R}^d$. This problem has a convex analytical Pareto front given by the following Pareto set:
\begin{equation*}
	0 \leq x_1 \leq 1 \text{ and } x_i=0 \text{ for } i \in \llbracket 2; d\rrbracket
\end{equation*}
The ZDT2 problem is given by the following functions \cite{charayron_towards_2023}:

\begin{equation*}
	\begin{array}{c}
	(J_1) \quad \left\{\begin{array}{l}
		f_{hf}(x) = x_1 \\[9pt]
		f_{lf}(x) = 0.8 x_1 + 0.2
	\end{array}\right. \qquad
	(J_2) \quad \left\{\begin{array}{l}
		f_{hf}(x) = u(x) v(x) \\[9pt]
		f_{lf}(x) = (0.9 u(x) + 0.2)(1.1 v(x) - 0.2) \\[9pt]
		u(x) = 1 + \frac{9}{d-1}\sum_{i=2}^{d} x_i \\[9pt]
		v(x) = 1 - \left(\frac{x_1}{u(x)}\right)^2
	\end{array}\right.
	\end{array}
\end{equation*}
It has a concave analytic Pareto front given by the same Pareto set as that of problem ZDT1. \par

\begin{figure}[!htb]
\centering
\includegraphics[width=.33\textwidth]{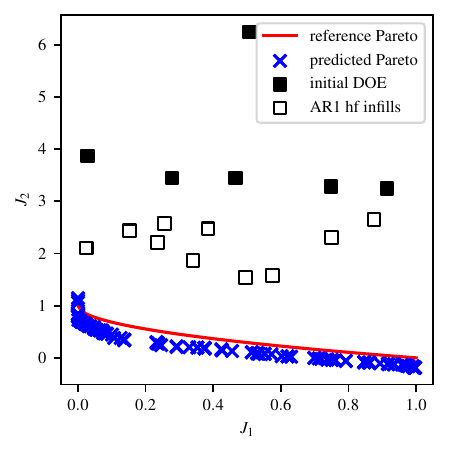}\hfill
\includegraphics[width=.33\textwidth]{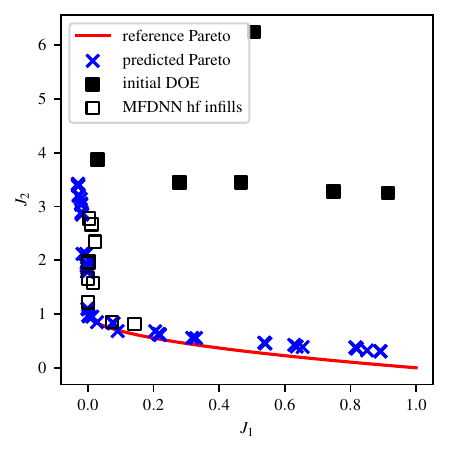}\hfill
\includegraphics[width=.33\textwidth]{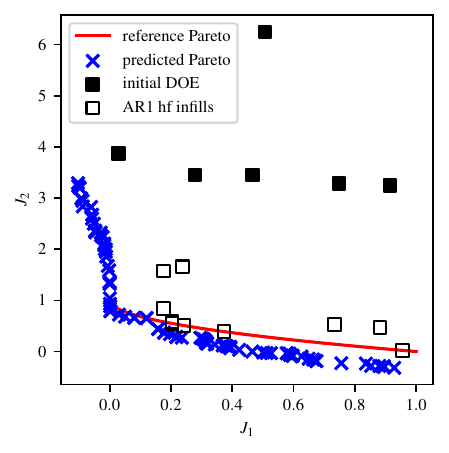}\\
\includegraphics[width=.33\textwidth]{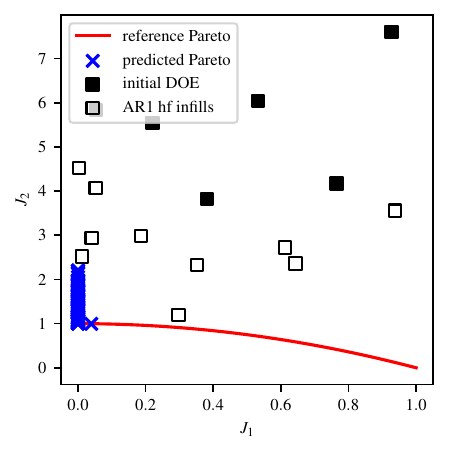}\hfill
\includegraphics[width=.33\textwidth]{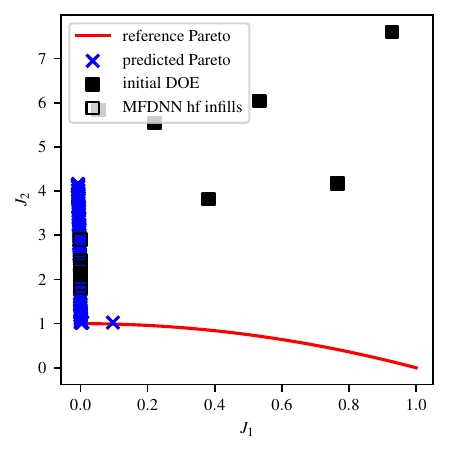}\hfill
\includegraphics[width=.33\textwidth]{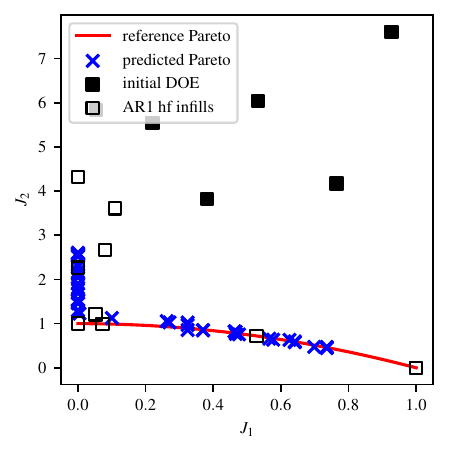}
\caption{One realization of the ten ZDT1 (top) and ZDT2 (bottom) optimizations for E1 (left), E2 (center) and E3 (right).}
\label{fig:zdt}
\end{figure}

Both problems are solved with their dimension $d$ set to 6 and the size of the initial low- and high-fidelity DOEs are respectively 12 and 6. The number of infill step $n_\text{iter}$ is 10 and at each step, a single high-fidelity and $n_{lf}=10$ low-fidelity samples are computed. \autoref{tab:infill} summarizes the results obtained with the optimization approaches introduced in \autoref{sec:mfo}. Only the IGD and IGD+ criteria are evaluated for the dataset Pareto and for each configuration, the results are averaged over ten realizations computed with distinct seeds. For each experiment, one realization is represented in \autoref{fig:zdt}. \par

\begin{table}[!htb]
	\centering
    \begin{tabular}{lc|cc|cc}
    \hline\\[-1.0em]
    Experiments & Infill strategy & \multicolumn{2}{c|}{ZDT1} & \multicolumn{2}{c}{ZDT2} \\
    & (see \autoref{sec:mfo}) & IGD & IGD+ & IGD & IGD+ \\
    \hline
    \hline\\[-1.0em]
        E1: AR1 & MPI, LCB \& ED & 1.20 (0.53) & 1.19 (0.53) & 1.89 (0.72) & 1.86 (0.74) \\
        E2: MFDNN  & NSGA-II \& ED & 0.85 (0.38) & 0.73 (0.42) & 0.69 (0.31) & 0.55 (0.29)  \\
        E3: AR1  & $\alpha_{WB2S, MPI}$, LCB \& ED & \textbf{0.12} (0.026) & \textbf{0.083} (0.018) & \textbf{0.25} (0.11) & \textbf{0.14} (0.086) \\
    \hline
    \end{tabular}
    \caption{Averaged value and standard deviation (between parenthesis) of IGD and IGD+ obtained for ten repetitions. The best performances are written in bold.}
    \label{tab:infill}
\end{table}

\autoref{tab:infill} shows that for both problems, the Bayesian strategy with the regularized infill criterion performs the best. Indeed, as illustrated in \autoref{fig:zdt}, although the non-regularized Bayesian strategy provides relevant infill samples, they are still distant to the reference Pareto front. This is inherent to the MPI acquisition function which does not quantify the infill improvement itself but only its probability to be an improvement with respect to the other high-fidelity samples. With regularization, the Bayesian strategy systematically manages to give infill samples on the Pareto front. On average, this approach improves the Pareto-front metrics by an order of magnitude with respect to its non-regularized counterpart. On the other hand, the non-Bayesian strategy yields results somewhere between the Bayesian strategies (see \autoref{tab:infill}). But, as evident from \autoref{fig:zdt}, the NSGA-II algorithm tends to vertically extend the Pareto front on the left-side of the domain (i.e. for $J_1\rightarrow0$) in both cases. Since the non-Bayesian approach fully relies on NSGA-II runs, this results in the computation of high-fidelity infill samples in this particular area and a poor overall coverage of the Pareto front. \par

Based on these preliminary results, we retain the AR1 model with regularized Bayesian infill strategy and the MFDNN model with non-Bayesian infill strategy as the two most promising approaches to be further investigated for a more realistic ASO problem.

\section{A multi-objective optimization problem: the DLR low Reynolds number OGV cascade}\label{sec:cascade}

In the following we apply the two proposed multi-fidelity and multi-objective strategies to the ASO of a low Reynolds number (LRN) linear cascade of outlet guiding vane (OGV) blades. The system has been extensively investigated at the German aerospace center (DLR) \cite{hergt_riblet_2015, Hergt2024vol15no4tp08}, and tested in the transonic cascade wind tunnel of DLR in Cologne. This use-case has been recently revived as part of the SciFi-Turbo\footnote{The project's website is available at the following link: \url{https://scifiturbo.eu/}.} European research project.

\subsection{Optimization problem description}
The LRN-OGV use case aims at simultaneously optimizing the cascade efficiency for three operating points: the aerodynamic design point (ADP), whose flow conditions are given in \autoref{tab:adp}, the operating point 1 (OP1) and the operating point 2 (OP2) with incidence angle variations of respectively +/-5$^\circ$ with respect to the ADP incidence angle. As such, these operating points are representative of near-stall and near-choking conditions. In addition, several geometric constraints are applied to make sure that the selected design can be manufactured. With these requirements, the optimization problem writes:
\begin{align}
	& \text{minimize } \left\{w_{ADP},\,  w_{OP} \right\}, \\
	\notag & w_{OP} = \frac{1}{2}\left(w_{OP1} + w_{OP2}\right),
\end{align}
where $w$ is the pressure loss coefficient defined as:
\begin{equation}
	w = \frac{P_{01} - P_{02}}{P_{01} - P_1},
\end{equation}
with $P_1$ the inlet static pressure, $P_{01}$ and $P_{02}$ the inlet and outlet pitch-wise averaged total pressures. The baseline relative design constraints are:
\begin{align}
	& \frac{\max(X_{th})}{c_{ax}} = \pm 20 \%, \\
	& \frac{\text{Area}}{c^2} = \pm 20 \%, \\
	& \frac{X_{cg}}{c_{ax}} = \pm 20 \%,
\end{align}
where $\max(X_{th})$, $c_{ax}$, Area, $c$ and $X_{cg}$ respectively denote the axial position of maximal thickness, the axial chord, the blade area, the chord and the axial position of the center of gravity. The absolute design constraints are:
\begin{align}
	& \frac{r_{le}}{c} > 0.5 \%, \\
	& \frac{r_{te}}{c} > 0.5 \%,
\end{align}
where $r_{le}$ and $r_{te}$ correspond to the leading and trailing edge radii. \par
Due to the simultaneous optimization of the performance for three operating conditions, the computational cost of the objective function evaluation amounts to the execution of three simulations for each new design.

\begin{table}[!htb]
    \centering
    \begin{tabular}{ll}
    \hline\\[-1.0em]
    Inlet Mach number & $M_1 = 0.60$  \\
    Inlet Reynolds number & $Re = 1.5\times10^5$ \\
    Inlet flow angle & $\beta=133^\circ$ \\
    Inlet total pressure & $P_{01}=18\,417$ Pa \\
    Outlet static pressure & $P_2=16\,258$ Pa \\
    Chord & $c=70$ mm \\
    \hline
    \end{tabular}
    \caption{LRN compressor cascade flow conditions at ADP.}
    \label{tab:adp}
\end{table}

\subsection{Baseline solutions}
In the multi-fidelity approach, two levels of fidelity are considered as follows. The low-fidelity model corresponds to RANS simulations run on a coarse mesh automatically generated using the open source \textit{gmsh} software\footnote{\textit{gmsh} is an open source mesh generation software hosted on gitlab (see \url{https://gmsh.info/}).}. The SA turbulence model available in the \textit{wolf} code is used to account for the turbulent stresses. The high-fidelity solutions, in turn, are obtained by solving the Navier--Stokes equations supplemented with the SA-BCM model on an anisotropic adapted mesh. \par

\begin{figure}[!htb]
\centering
\begin{minipage}{.49\textwidth}
  \centering
  \includegraphics[width=\linewidth]{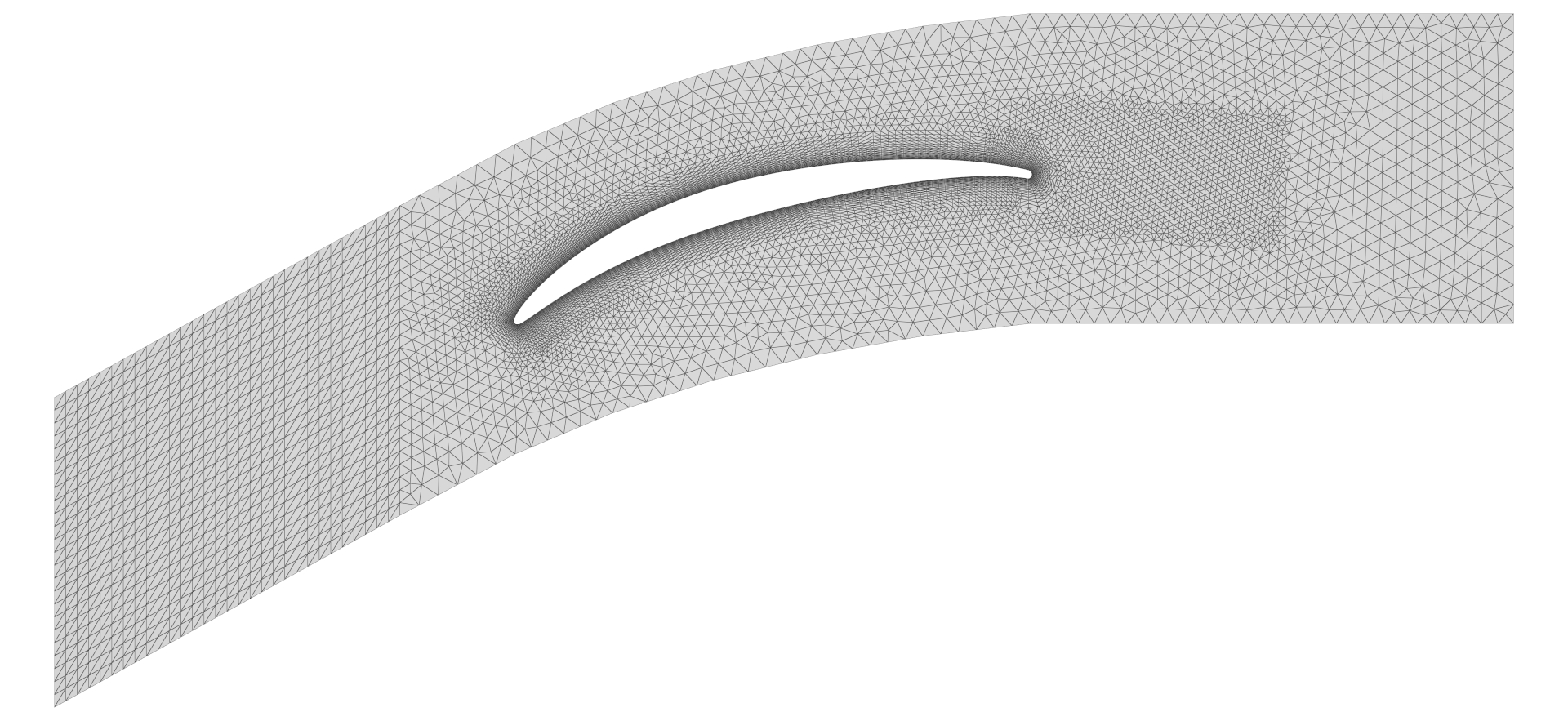}
  \caption{LRN-OGV cascade: coarse mesh composed of $\mathcal{N}_e=16\,000$ elements.}
  \label{fig:lrn-cascade-mesh}
\end{minipage}\hfill
\begin{minipage}{.49\textwidth}
  \centering
  \includegraphics[width=\linewidth]{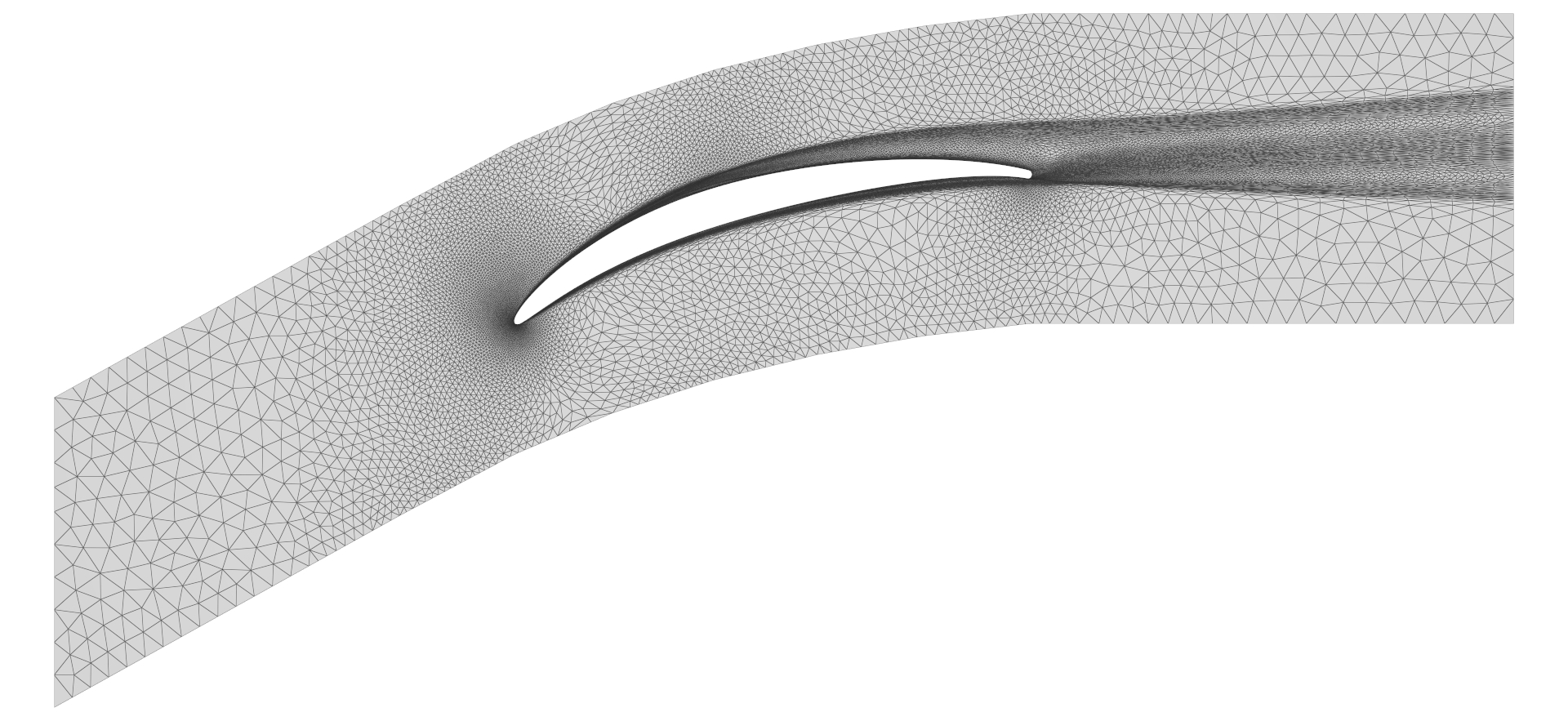}
  \caption{LRN-OGV cascade: adapted mesh (based on the SA-BCM model) of complexity $\mathcal{N}_c=16\,000$ and composed of $\mathcal{N}_e=38\,000$ cells.}
  \label{fig:lrn-cascade-adapted-mesh}
\end{minipage}
\end{figure}

\begin{figure}[!htb]
\centering
\begin{minipage}[t]{.49\textwidth}
  \centering
  \includegraphics[width=\linewidth]{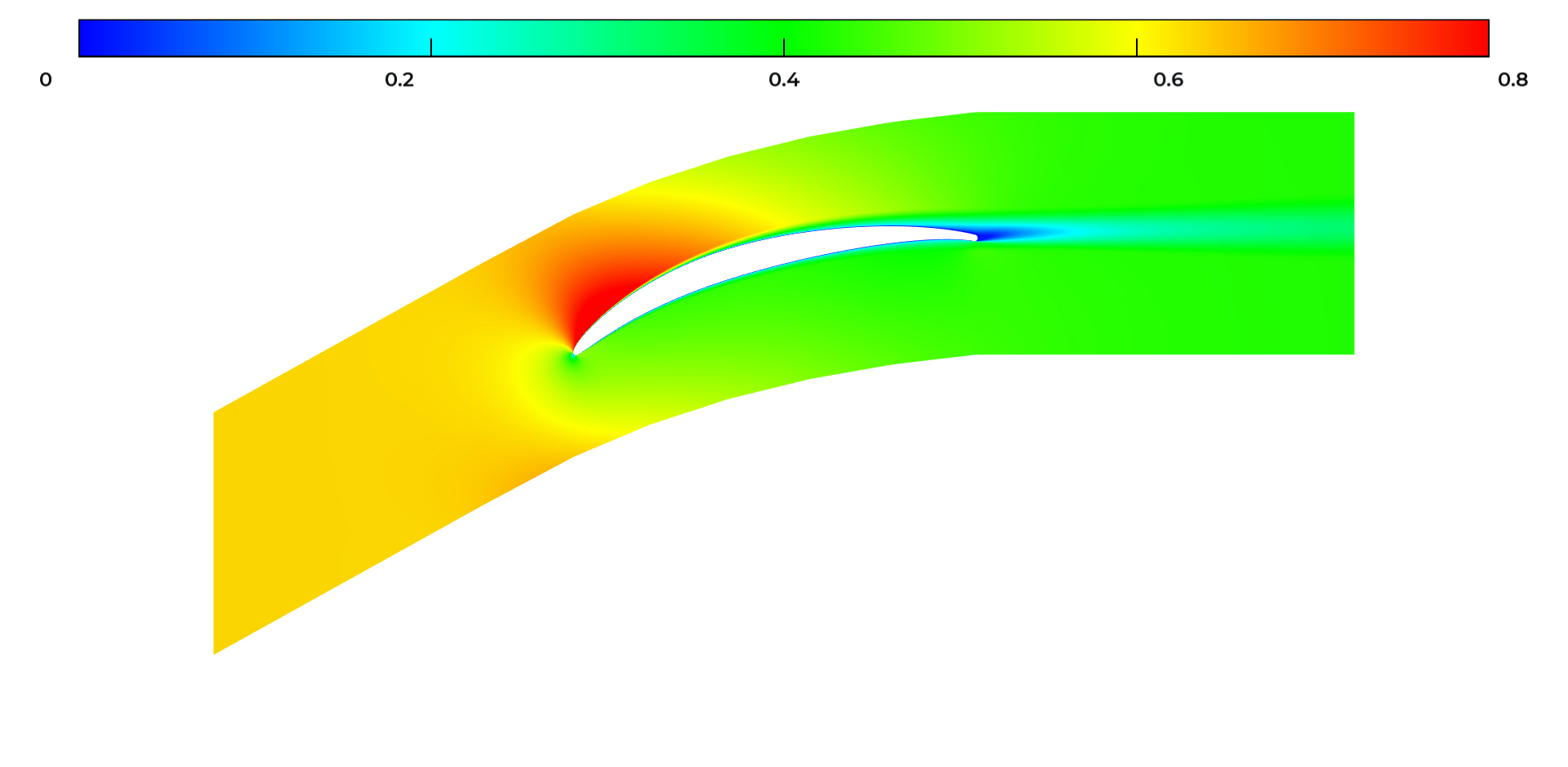}
  \caption{LRN-OGV cascade: Mach field for the RANS SA model and  coarse mesh.}
  \label{fig:lrn-cascade-mach}
\end{minipage}\hfill
\begin{minipage}[t]{.49\textwidth}
  \centering
  \includegraphics[width=\linewidth]{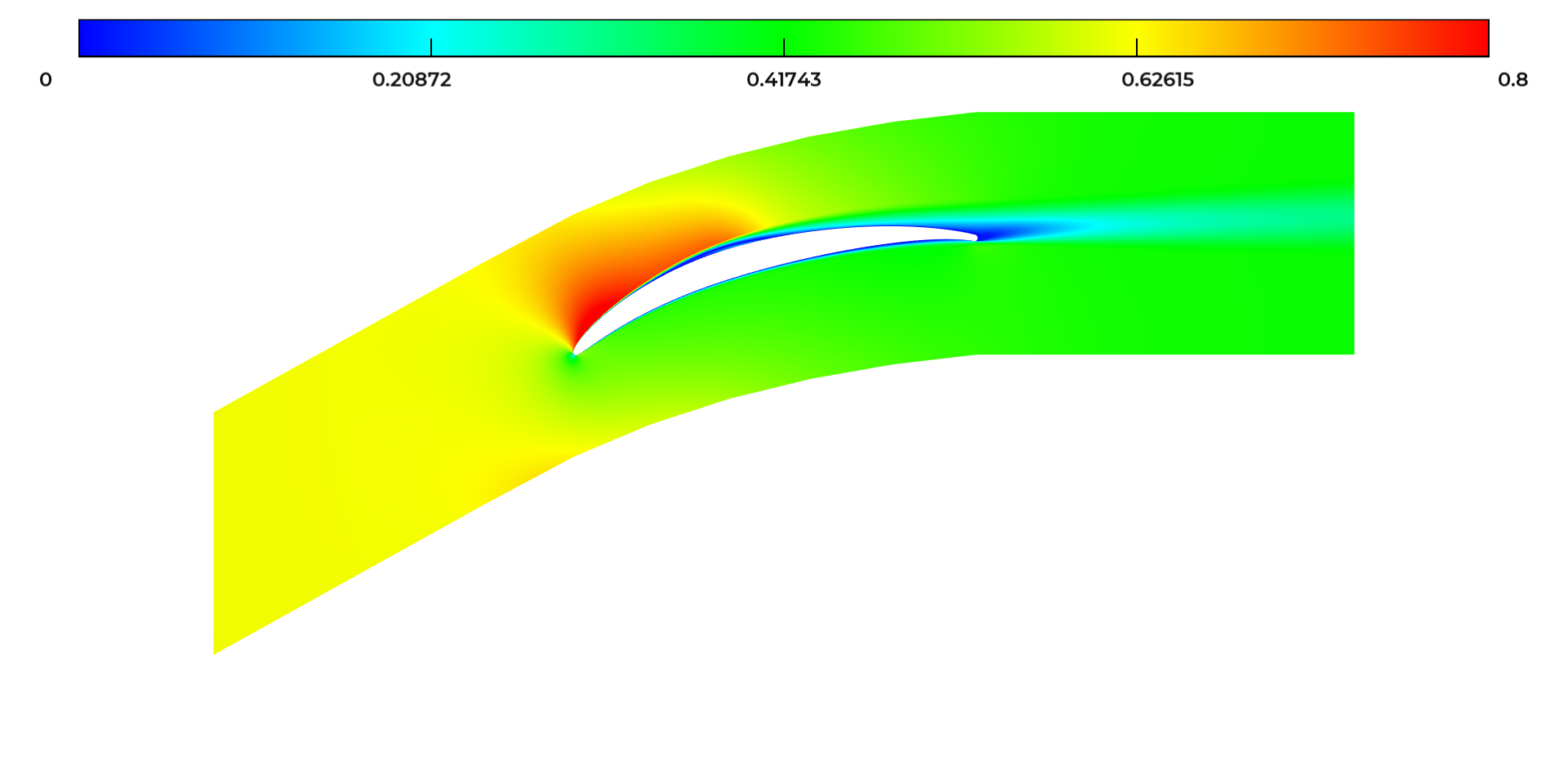}
  \caption{LRN-OGV cascade: Mach field for the  RANS SA-BCM model and adapted mesh.}
  \label{fig:lrn-cascade-adapted-mach}
\end{minipage}
\end{figure}

\begin{figure}[!htb]
  \centering
  \includegraphics[width=0.5\linewidth]{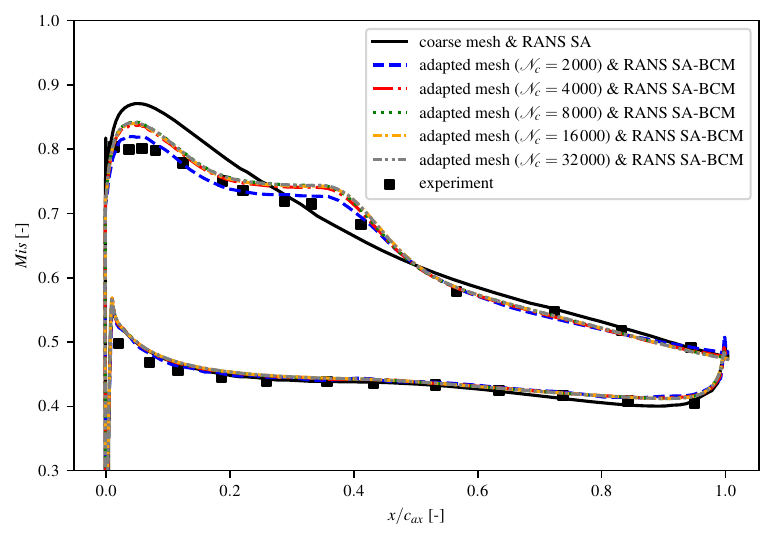}
  \caption{Distributions of the isentropic Mach number along the blade wall for low- and high-fidelity solvers and experimental data from \cite{hergt_riblet_2015}. The high-fidelity model results are reported for various mesh complexities.}
  \label{fig:lrn-cascade-mis}
\end{figure}

For the baseline ADP conditions, the corresponding low- and high-fidelity meshes are shown in \autoref{fig:lrn-cascade-mesh} and \ref{fig:lrn-cascade-adapted-mesh}. The corresponding Mach fields are presented in \autoref{fig:lrn-cascade-mach} and \ref{fig:lrn-cascade-adapted-mach}. The flow is characterized by a laminar separation bubble at the suction side, which is captured by the high-fidelity model but is missed by the low-fidelity one. The bubble strongly affects the distribution of the isentropic Mach number $M_{is}=\left[\frac{2}{\gamma-1}\left((\frac{P_{01}}{P})^{\frac{\gamma-1}{\gamma}}-1 \right)\right]^\frac{1}{2}$ (with $P$ the local static pressure and $\gamma$ the gas isentropic exponent) along the wall, as reported in \autoref{fig:lrn-cascade-mis} where the results are compared with the measurements obtained in DLR testing facility. The complexity of the adapted grid is progressively increased until achieving grid convergence. A complexity $\mathcal{N}_c=16\,000$ is finally deemed sufficient to obtained a mesh-converged solution. \par

In \autoref{tab:losscoef} we report the loss coefficients predicted by the low- and high-fidelity models. Due to the parallelism-induced non-deterministic nature of the adaptation procedure, 10 simulations are performed to estimate the uncertainty in the predicted loss coefficients. For ADP and OP2, the standard deviation is less than 1\% of the mean value, while it is slightly below 2\% for the near-stall condition OP1, for which the SA-BCM model tends to predict unsteady flow due to the appearance of a large separation bubble.

\begin{table}[!htb]
    \centering
    \begin{tabular}{ccccccc}
    \hline\\[-1.0em]
    Fidelity & Turbulence model & Mesh ($\mathcal{N}_e$) & $w_{ADP}$ (std$\%$) & $w_{OP1}$ (std$\%$) & $w_{OP2}$ (std$\%$) & $w_{OP}$ (std$\%$)  \\
    \hline
    \hline\\[-1.0em]
	Low & RANS SA & coarse ($16\,000$) & 0.0360 (-) & 0.0524 (-) & 0.0411 (-) & 0.0467 (-) \\
	High & RANS SA-BCM & adapted ($38\,000$) & 0.0403 (0.6) & 0.0929 (1.8) & 0.0430 (0.3) & 0.0679 (1.2) \\
    \hline
    \end{tabular}
    \caption{LRN compressor cascade baseline loss coefficients for the low- and high-fidelity solutions. The high-fidelity results are averaged over 10 executions and std\% indicates the standard deviation relative to the mean value in \%.}
    \label{tab:losscoef}
\end{table}

\subsection{CFD-based single-fidelity optimizations}
To investigate the effect of the CFD model on the optimization results, we first run a low- and a high-fidelity optimization without the assistance of a surrogate model using the NSGA-II optimizer with 50 generations of 20 candidates. The turn-around time of the high-fidelity optimization is approximately 2 days on a 48 cores/96 threads workstation and about 3 hours for the low-fidelity optimization. \par

\begin{figure}[!htb]
\centering
\begin{subfigure}[b]{0.49\textwidth}
	\centering
  	\includegraphics[width=\linewidth]{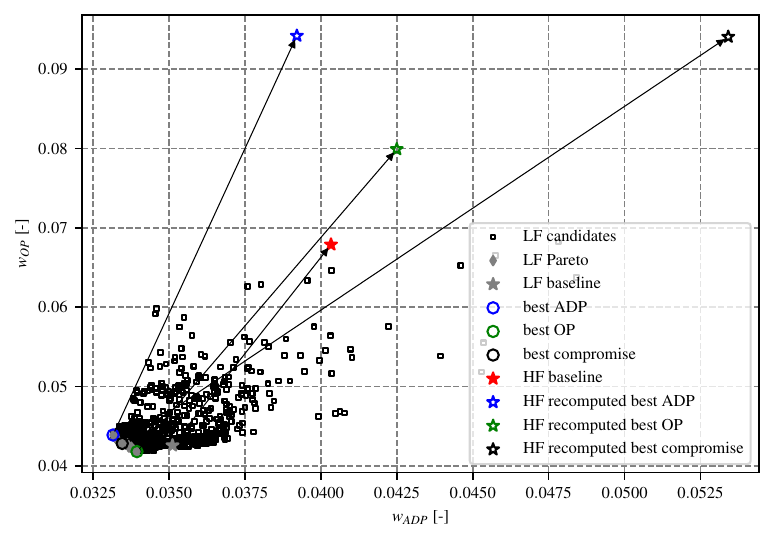}
\end{subfigure} \hfill
\begin{subfigure}[b]{0.49\textwidth}
	\centering
  	\includegraphics[width=\linewidth]{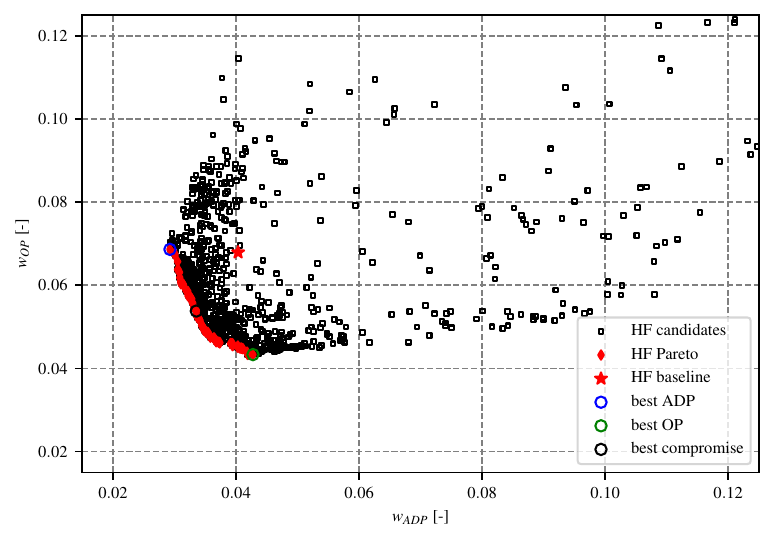}
\end{subfigure}
  \caption{Single-fidelity CFD-based Pareto fronts for the low- (left) and high-fidelity (right) models. The Pareto front designs are represented in plain diamonds; the best candidates with respect to $w_{ADP}$ and $w_{OP}$ are circled in blue and green; the best compromise at the center of the Pareto front is circled in black.}
  \label{fig:ref}
\end{figure}

The results are illustrated in \autoref{fig:ref} where the final Pareto fronts after 50 generations are represented with plain diamonds, and the optimal solutions and the best compromise between the two objectives are circled. The low- and high-fidelity  Pareto fronts show dramatic differences, arguing for the use of a high-fidelity model to drive the optimization. The recomputation of the optimal low-fidelity candidates with the high-fidelity model highlights the complexity of the relationship between both fidelities. Indeed, the correlation is opposite to that of the LES/RANS correlation of \cite{matar_camille_analysis_2024, matar2025cost} where the low-fidelity model systematically over-predicted the high-fidelity objective function values. This makes the low fidelity model particularly misleading to the optimization if not corrected with high-fidelity information. \par

\begin{figure}[!htb]
\centering
  \includegraphics[width=0.8\linewidth]{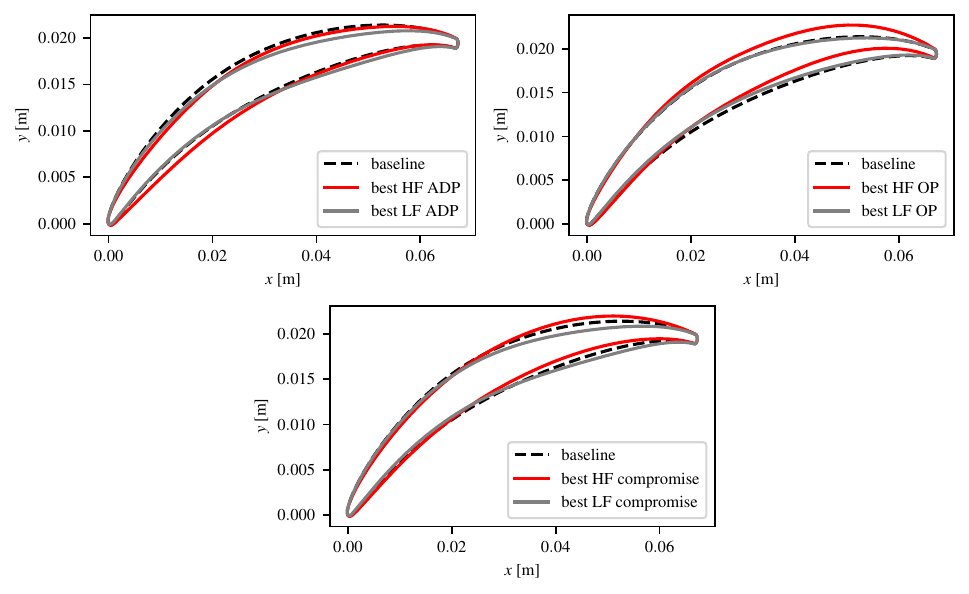}
  \caption{Best-performing profiles with respect to $w_{ADP}$ (top left), $w_{OP}$ (top right) and best compromise (bottom).}
  \label{fig:blade}
\end{figure}

The shape of the Pareto fronts hints that the Pareto-optimal high-fidelity candidates encompass very different geometries,  whereas the tight low-fidelity Pareto front and small variations in $w_{OP}$ suggest more similar shapes. This is confirmed by inspection of \autoref{fig:blade}, which shows the optimal blade profiles for the two models. According to the high-fidelity model, the best-performing profile at nominal condition exhibits a flatter pressure side than the baseline, whereas the low-fidelity best individual mostly differ from the baseline at the rear of the suction side. Furthermore, the profiles ensuring the best off-design performance are highly cambered in the rear part and differ significantly from the baseline, which is not the case for the low-fidelity best-performing individual with respect to this criterion. Similarly, significant differences are also observed for the designs corresponding to the best compromise between the two objectives.

\subsection{Surrogate-assisted multi-fidelity optimization}
The LRN-OGV cascade optimization problem is then solved using the Bayesian and non-Bayesian multi-fidelity surrogates with six configurations detailed in \autoref{tab:exp}. The MFDNN architecture is analogous to that of Zhang et al. \cite{zhang_multi-fidelity_2021} as given in \autoref{tab:mfdnn} where the dimensions of the output layers $y_L$, $y_H$ and the parameter $\beta$ are extended to the number of objectives.  Regarding the initial DOEs, they are nested and made of $n-1$  LHS samples and the baseline results. This seems determinant as it minimizes the risks of having high-fidelity infill samples with poorer performances. In order to perform fair comparisons between each configuration, the same initial DOEs are used for the non-reduced optimizations (E1, E3, E5). In the same way, the initial DOEs of the two POD-based optimizations (E2, E4, E6) are the same and built by inverse reconstruction of the reduced parameters corresponding to the non-reduced configurations (see \ref{app:pod}). For each experience, five runs are performed.  \par

\begin{table}[!htb]
	\centering
    \begin{tabular}{lc|cc|cc|r}
    \hline\\[-1.0em]
    Experiments & Problem & \multicolumn{2}{c|}{DOEs} & \multicolumn{2}{c|}{Total infills} & Infill strategy \\
    & dimension & LF & HF & LF & HF & (see \autoref{sec:mfo}) \\
    \hline
    \hline\\[-1.0em]
        \rowcolor{light-light-gray}
        E1: AR1 & 8 & 99 + 1 & 9 + 1 & 100 & 10 & MPI, LCB \& ED \\
        E2: AR1 \& POD & 4 & 99 + 1 & 9 + 1 & 100 & 10 & MPI, LCB \& ED \\
        \rowcolor{light-light-gray}
        E3: MFDNN & 8 & 99 + 1 & 9 + 1 & 100 & 10 & NSGA-II \& ED \\
        E4: MFDNN \& POD & 4 & 99 + 1 & 9 + 1 & 100 & 10 & NSGA-II \& ED \\
        \rowcolor{light-light-gray}
        E5: AR1 & 8 & 99 + 1 & 9 + 1 & 100 & 10 & $\alpha_{WB2S, MPI}$, LCB \& ED \\
        E6: AR1 \& POD & 4 & 99 + 1 & 9 + 1 & 100 & 10 & $\alpha_{WB2S, MPI}$, LCB \& ED \\
    \hline
    \end{tabular}
    \caption{Multi-fidelity multi-objective assisted optimization configurations. The "Total infills" column indicates the total number of infill samples of each fidelity. The number of infill steps is equal to the number of HF infills.}
    \label{tab:exp}
\end{table}

In the end, each experiment yields two Pareto fronts. The dataset Pareto which corresponds to the one formed by the high-fidelity points, and the predicted Pareto which is given by the post-processing NSGA-II execution coupled with the final version of the surrogate model.

\subsubsection*{Dataset Pareto metrics evaluation}
We begin with the analysis of the Pareto-front metrics. \autoref{tab:res} summarizes the optimization results with respect to the performance metrics introduced in \autoref{sec:metrics}. As evident from the first column of each block, it appears that in spite of our efforts to reduce the discrepancies between the initial DOEs, the cumulative effect of the POD reconstruction error and the model error in case of strong boundary layer detachment result in non-negligible objective value differences. For this reason, the POD-based configurations depart with a slightly poorer initial high-fidelity DOE in view of all performances metrics. \par

\begin{table}[!htb]
	\centering
    \begin{tabular}{l|ccc|ccc|ccc}
    \hline\\[-1.0em]
    Exp. & \multicolumn{3}{|c}{IGD ($\downarrow$)} & \multicolumn{3}{|c}{IGD+ ($\downarrow$)} & \multicolumn{3}{|c}{HV ($\uparrow$)} \\
    & init. & final & ratio & init. & final & ratio & init. & final & ratio \\
    \hline
    \hline\\[-1.0em]
    		Reference & - & 0 & - & - & 0 & - & - & $1.7\times10^{-4}$ & - \\
    	\hline\\[-1.0em]
    		\multicolumn{10}{c}{Performance metrics after 5 infill steps} \\
    	\hline\\[-1.0em]
        \rowcolor{light-light-gray}
        E1 & $8.8\times10^{-3}$ & $\mathbf{6.8\times10^{-3}}$ & $\mathbf{1.3}$ & $8.1\times10^{-3}$ & $\mathbf{5.7\times10^{-3}}$ & $\mathbf{1.5}$ & $1.3\times10^{-5}$ & $2.1\times10^{-5}$ & 1.6 \\
        E2 & $9.0\times10^{-3}$ & $7.1\times10^{-3}$ & $1.3$ & $8.9\times10^{-3}$ & $6.9\times10^{-3}$ & $1.4$ & $3.9\times10^{-6}$ & $\mathbf{3.8\times10^{-5}}$ & $\mathbf{13.0}$ \\
        \rowcolor{light-light-gray}
        E3 & $8.8\times10^{-3}$ & $8.0\times10^{-3}$ & $1.1$ & $8.1\times10^{-3}$ & $7.5\times10^{-3}$ & $1.1$ & $1.3\times10^{-5}$ & $1.8\times10^{-5}$ & 1.4 \\
        E4 & $9.0\times10^{-3}$ & $8.3\times10^{-3}$ & $1.1$ & $8.9\times10^{-3}$ & $7.7\times10^{-3}$ & $1.3$ & $3.9\times10^{-6}$ & $1.9\times10^{-5}$ & 5.0 \\
        \rowcolor{light-light-gray}
        E5 & $8.8\times10^{-3}$ & $8.6\times10^{-3}$ & 1.0 & $8.1\times10^{-3}$ & $7.4\times10^{-3}$ & 1.2 & $1.3\times10^{-5}$ & $1.6\times10^{-5}$ & 1.2 \\
        E6 & $9.0\times10^{-3}$ & $9.0 \times 10^{-3}$ & 1.0 & $8.9\times10^{-3}$ & $8.9\times10^{-3}$ & 1.0 & $3.9\times10^{-6}$ & $3.9\times10^{-6}$ & 1.0 \\
      \hline\\[-1.0em]
    		\multicolumn{10}{c}{Performance metrics after 10 infill steps} \\
    	\hline\\[-1.0em]
        \rowcolor{light-light-gray}
        E1 & $8.8\times10^{-3}$ & $5.2\times10^{-3}$ & $1.8$ & $8.1\times10^{-3}$ & $4.0\times10^{-3}$ & $2.1$ & $1.3\times10^{-5}$ & $5.6\times10^{-5}$ & 4.4 \\
        E2 & $9.0\times10^{-3}$ & $\mathbf{4.9\times10^{-3}}$ & $\mathbf{1.9}$ & $8.9\times10^{-3}$ & $4.4\times10^{-3}$ & $2.1$ & $3.9\times10^{-6}$ & $\mathbf{8.4\times10^{-5}}$ & $\mathbf{31.7}$ \\
        \rowcolor{light-light-gray}
        E3 & $8.8\times10^{-3}$ & $7.2\times10^{-3}$ & $1.3$ & $8.1\times10^{-3}$ & $5.7\times10^{-3}$ & $1.5$ & $1.3\times10^{-5}$ & $2.6\times10^{-5}$ & 2.1 \\
        E4 & $9.0\times10^{-3}$ & $6.0\times10^{-3}$ & $1.6$ & $8.9\times10^{-3}$ & $5.6\times10^{-3}$ & $1.7$ & $3.9\times10^{-6}$ & $3.6\times10^{-5}$ & 10.3 \\
        \rowcolor{light-light-gray}
        E5 & $8.8\times10^{-3}$ & $7.5\times10^{-3}$ & 1.2 & $8.1\times10^{-3}$ & $6.7\times10^{-3}$ & 1.3 & $1.3\times10^{-5}$ & $1.6\times10^{-5}$ & 1.2 \\
        E6 & $9.\times10^{-3}$ & $5.0\times10^{-3}$ & 1.8 & $8.9\times10^{-3}$ & $\mathbf{3.6\times10^{-3}}$ & $\mathbf{2.5}$ & $3.9\times10^{-6}$ & $7.0\times10^{-5}$ & 23.3 \\
    \hline
    \end{tabular}
    \caption{Comparison of the dataset Pareto mean IGD, IGD+ and HV for the different configurations after 5 and 10 infill steps over five runs. For each metric, "$\downarrow$" indicates the lower the better, "$\uparrow$" the opposite. The ratio is always to be maximized and in each block, the best performance is written in bold.}
    \label{tab:res}
\end{table}

To get a grasp of the optimization speed, the metrics are evaluated after five and ten infill steps. After five infills, it seems that both POD-coupled models (E2, E4) offer a better exploration of the design space with a significant increase in the hypervolume below the baseline objective values. Regarding the inverse generational distance criteria, the AR1 model without data reduction (E1) yields the best IGD and IGD+ values and ratios but very close to that of the POD-coupled AR1 (E2). Given the problem sensitivity and in view of the three metrics, it appears that for a very low number of infills, the AR1-based configurations (E1, E2) are the most efficient. \par

At the end of the ten infills, the results indicate that the Bayesian-based strategies (E1, E2, E6) compare well and the POD-coupled versions (E2, E6) provide the best performances with respect to the three metrics. On average, the non-Bayesian strategy with or without POD-coupling (E3, E4) exhibit poorer performances than their Bayesian counterparts (E1, E2, E6). \par

Regarding the use of a regularized infill criterion, the results fail to show significant improvements. Regardless of the problem dimension (E5, E6), the regularization does not improve the optimization speed as all metrics ratio remain close to 1 after five infill steps. After ten infill steps however, the POD-coupled version (E6) performs as good as the non-regularized Bayesian strategy while without reduction (E5), the performances are surprisingly slightly poorer as those obtained with the non-Bayesian strategy (E3). This could mean that the efficiency of the regularization is highly dependent on the problem's dimensions. \par

\subsubsection*{Predicted Pareto evaluation}
\begin{figure}[!htb]
\centering
    \begin{subfigure}[b]{0.49\textwidth}
	\centering
  	\includegraphics[width=\linewidth]{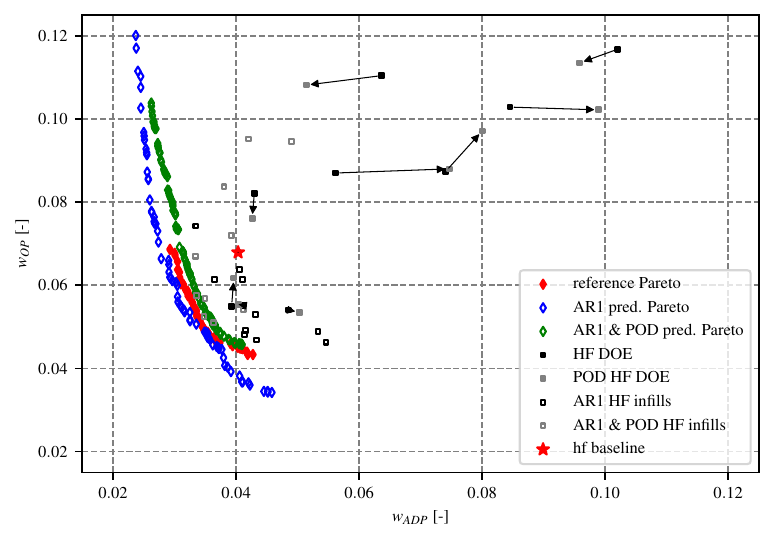}
\end{subfigure} \hfill
\begin{subfigure}[b]{0.49\textwidth}
	\centering
  	\includegraphics[width=\linewidth]{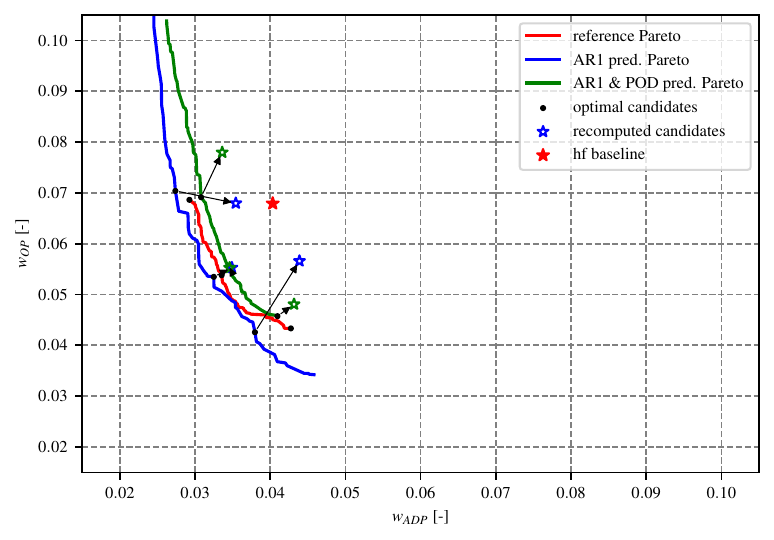}
\end{subfigure}
  \caption{Comparison of the AR1-based models optimization results (E1, E2) for one of the five runs. The predicted Pareto fronts are represented with blue (E1) and green (E2) empty diamonds. On the left side, the arrows between plain squares indicate the cumulative effect of the model and POD reconstruction errors between the initial high-fidelity DOEs. On the right side, the arrow between the empty diamonds and stars indicate the error between the actual and predicted values.}
  \label{fig:opt-ar1}
\end{figure}

\begin{figure}[!htb]
\centering
    \begin{subfigure}[b]{0.49\textwidth}
	\centering
  	\includegraphics[width=\linewidth]{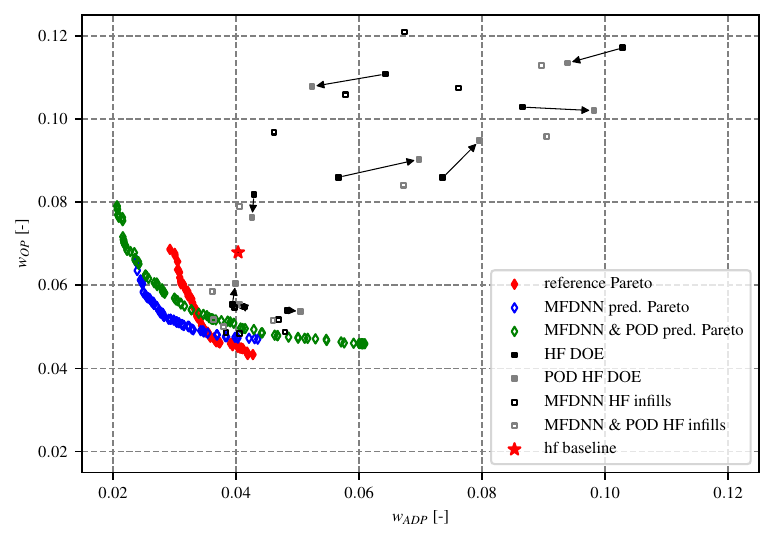}
\end{subfigure} \hfill
\begin{subfigure}[b]{0.49\textwidth}
	\centering
  	\includegraphics[width=\linewidth]{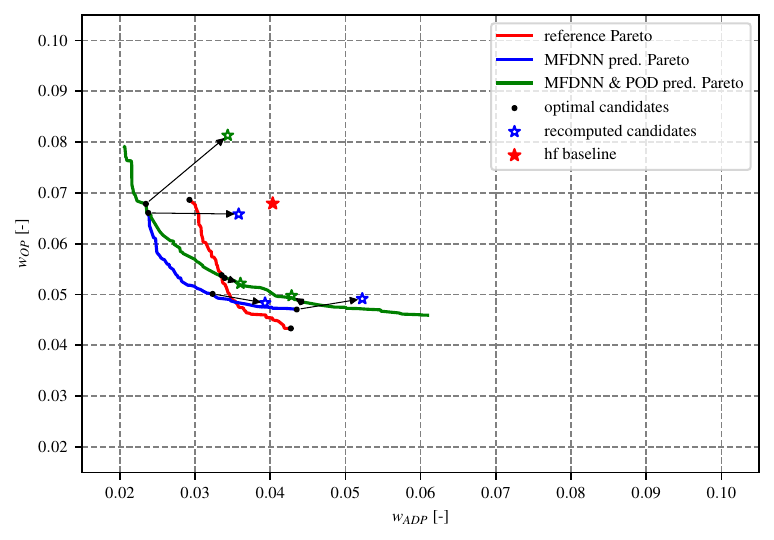}
\end{subfigure}
  \caption{Comparison of the MFDNN-based models optimization (E3, E4) results for one of the five runs. The predicted Pareto fronts are represented with blue (E3) and green (E4) empty diamonds. On the left side, the arrows between plain squares indicate the cumulative effect of the model and POD reconstruction errors between the initial high-fidelity DOEs. On the right side, the arrow between the empty diamonds and stars indicate the error between the actual and predicted values.}
  \label{fig:opt-mfdnn}
\end{figure}

\begin{figure}[!htb]
\centering
  \begin{subfigure}[b]{0.49\textwidth}
	\centering
  	\includegraphics[width=\linewidth]{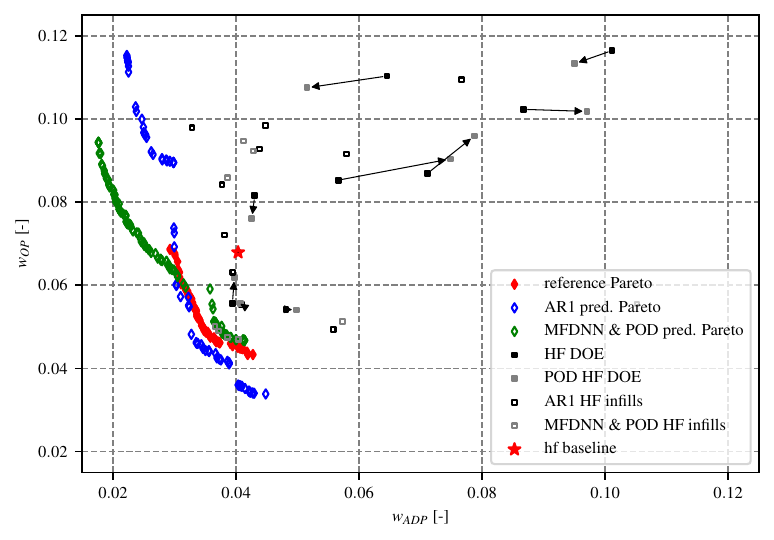}
\end{subfigure} \hfill
\begin{subfigure}[b]{0.49\textwidth}
	\centering
  	\includegraphics[width=\linewidth]{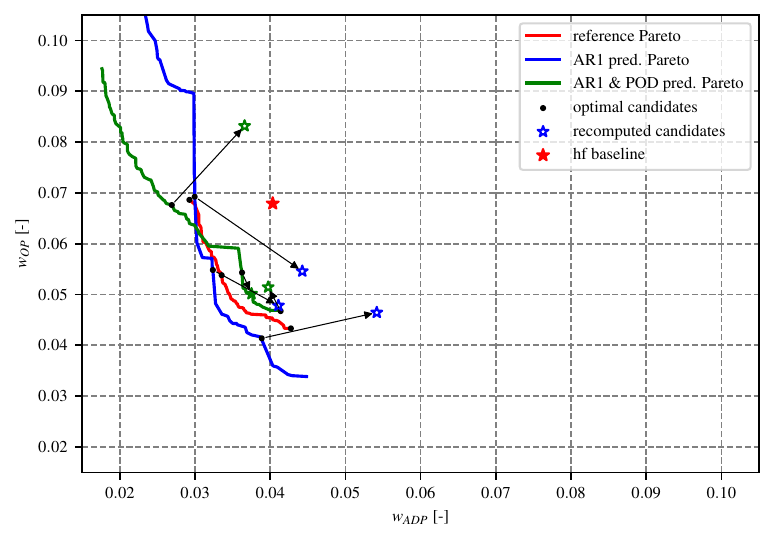}
\end{subfigure}
  \caption{Comparison of the AR1-based models and the regularized infill optimization (E5, E6) results for one of the five runs. The predicted Pareto fronts are represented with blue (E5) and green (E6) empty diamonds. On the left side, the arrows between plain squares indicate the cumulative effect of the model and POD reconstruction errors between the initial high-fidelity DOEs. On the right side, the arrow between the empty diamonds and stars indicate the error between the actual and predicted values.}
  \label{fig:opt-reg-ar1}
\end{figure}

Finally for each strategy, examples of predicted Pareto obtained with a post-processing NSGA-II assisted execution are represented in \autoref{fig:opt-ar1}, \ref{fig:opt-mfdnn} and \autoref{fig:opt-reg-ar1}. As explained in \cite{charayron_towards_2023}, this last step can result in an unreliable predicted Pareto front if the model is not accurate enough. This word of caution is illustrated in this example where the predicted Pareto fronts extend in physically irrelevant areas of the design space. To assess the reliability of the predicted Pareto, the predicted candidates with the objective values the closest to those of the reference candidates (see \autoref{fig:ref}) are extracted from the Pareto sets and recomputed with the high-fidelity model. \par

\begin{figure}[!htb]
\centering
  \includegraphics[width=0.8\linewidth]{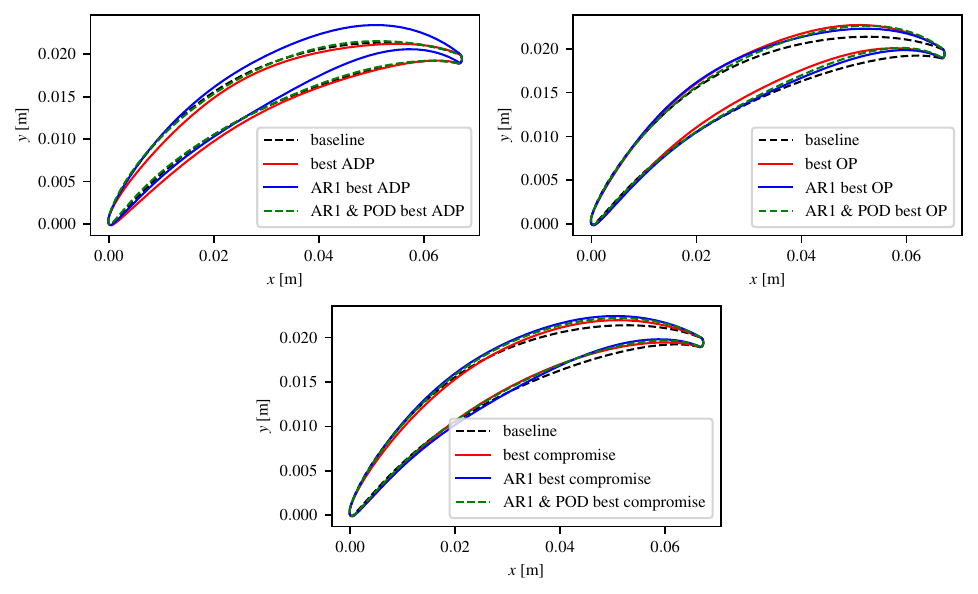}
  \caption{Comparison of the optimal predicted profiles for one of the five runs of the AR1-based model and regularized infill optimization (E5, E6). The profiles with the lowest $w_{ADP}$, $w_{OP}$ and at the center of the Pareto front are represented on the top left, top right and bottom.}
  \label{fig:pro-reg-ar1}
\end{figure}

As expected, the results indicate for all strategies that the region of the predicted Pareto which are not supported by the nearby presence of high-fidelity samples are highly unreliable. On the contrary, in the vicinity of the reference Pareto center which is more densely populated for all POD-coupled strategies (E2, E4, E6), the discrepancy between the predicted and the recomputed values are much smaller. To get an idea of how this observation translate into the design space, the profiles of the recomputed candidates of \autoref{fig:opt-reg-ar1} are given in \autoref{fig:pro-reg-ar1}. For the optimal candidates without POD-coupling (E5), the failure to compute high-fidelity samples near the reference Pareto results in highly inconsistent geometries with respect to the quantities of interest they are supposed to optimize. For the POD-coupled configuration (E6), the lack of high-fidelity samples in the low $w_{ADP}$ area of the objective spaces results in a blade somewhere between the reference optimal and the baseline geometries which leads to a smaller improvement of $w_{ADP}$ and a higher degradation of $w_{OP}$ than predicted. Regarding the other two optimal geometries, the agreement with the reference solution is much better.

\subsection*{Dataset Pareto evaluation}
Going back to the initial objective which is to improve the performances of the baseline cascade, \autoref{tab:perf} summarizes the average improvements resulting from the high-fidelity samples of each configuration. In comparison to the reference results, all strategies prove to work remarkably well and bring noticeable improvements with respect to both quantities and especially for $w_{OP}$. The slighter difficulty to improve $w_{ADP}$ comes from the fact that all five runs use the same initial DOEs which drive the optimization in the direction of better $w_{OP}$ due to the presence of initial high-fidelity samples in that area. The quality of the candidates with the optimal objectives trade-off (i.e. located at the center of the dataset Pareto) is more variable as two (E3, E5) out of the six strategies fail to provide a compromise candidate concurrently improving both quantities of interest. Nevertheless, all POD-coupled strategies (E2, E4, E6) are highly satisfactory.

\begin{table}[!htb]
	\centering
    \begin{tabular}{lccc}
    \hline\\[-1.0em]
    Exp. & opt. $w_{ADP}$ & opt. $w_{OP}$ & opt. trade-off \\
    & & & $w_{ADP}, w_{OP}$ \\
    \hline
    \hline\\[-1.0em]
    		Reference & 0.0293  & 0.0427  & $0.0335,\, 0.0538$ \\
    		& (27\%) & (36.2\%) & (16.8, 20.8) \\
    	\hline\\[-1.0em]
        \rowcolor{light-light-gray}
		E1 & 0.0357 & \textbf{0.0459} & $0.0373,\, 0.0516$ \\
        \rowcolor{light-light-gray}
		& (11.6) & \textbf{(32.4)} & (7.4, 24.1) \\
		E2 & \textbf{0.0334} & 0.0512 & $\mathbf{0.0357,\, 0.0530}$ \\
		& \textbf{(17.1)} & (24.6) & \textbf{(11.6, 21.9)} \\
        \rowcolor{light-light-gray}
		E3 & 0.0375 & 0.0484 & $0.0390,\, 0.0521$ \\
        \rowcolor{light-light-gray}
		& (7.0) & (28.7) & (3.2, 23.2) \\
		E4 & 0.0363 & 0.0485 & $0.0384,\, 0.0526$ \\
		& (10.0) & (28.6) & (4.7, 22.5) \\
        \rowcolor{light-light-gray}
		E5 & 0.0369 & 0.0472 & $0.0392,\, 0.0547$ \\
        \rowcolor{light-light-gray}
		& (8.6) & (30.4) & (2.8, 19.5) \\
		E6 & 0.0361 & 0.0476 & $\mathbf{0.0363,\, 0.0526}$ \\
		& (10.6) & (29.9) & \textbf{(10.0, 22.5)} \\
    \hline
    \end{tabular}
    \caption{Comparison of the average performances for the different configurations over five runs. The percentage of improvement with respect to the baseline values is given between parenthesis and the best performance is written in bold.}
    \label{tab:perf}
\end{table}

\subsection*{Summary}
In regards of the optimization metrics (IGD, IGD+ and HV), the results indicate that on average, the AR1 model associated to the considered Bayesian infill strategies (E1, E2, E6) performs better than its non-Bayesian counterparts (E3, E4). Regarding the extra NSGA-II step, although it does provide a denser Pareto front estimate, its extrapolation capacities are highly dependent on the nearby presence of high-fidelity samples. Finally, for all configurations, the dimension reduction through POD-coupling proved to significantly improve the quality of the dataset Pareto in regards of both the optimization metrics and the optimized quantities of interest. \par

\section{Conclusion}\label{sec:concl}
In the present work, the combination of four means to address the main challenges of ASO are considered under the assumption of extreme evaluation cost imbalance and strong budget constraint. In that respect, mesh adaptation, shape parametrization data reduction, multi-fidelity surrogate and adaptive infill are leveraged to solve the optimization under constraint of a low Reynolds number compressor cascade designed at DLR \cite{hergt_riblet_2015, Hergt2024vol15no4tp08}. The blade shape is parametrized with eight FFD control points and three optimization approaches are considered. Bayesian infill (with or without regularization) strategies are associated to AR1 models and a non-Bayesian infill strategy (based on NSGA-II) is associated to MFDNN models. In addition, data reduction is performed with POD to halve the number of design variables hence yielding a total of six optimization configurations. For all configurations, the multi-fidelity models were trained by combining low- and high-fidelity data respectively obtained with the RANS SA model for coarse meshes and the transition RANS SA-BCM model for error-free adapted meshes. At the end of the adaptive optimization, a post-processing step which consists of running NSGA-II assisted with the final version of the surrogate model was performed as suggested in \cite{charayron_towards_2023}.   \par

For the studied problem, the results showed that, in regards of the considered optimization metrics (IGD, IGD+ and HV), the AR1 model associated to Bayesian approaches perform better than the MFDNN model and its non-Bayesian strategy, namely in the very few high-fidelity sample regime targeted in the study. Considering the benefits of the NSGA-II post-processing step, the results showed that this contributes a denser Pareto, its reliability is highly dependent on the nearby presence of high-fidelity samples. It is then advised to recompute the high-fidelity solution of any geometry extracted from the predicted Pareto. Finally, for all approaches, dimension reduction with POD proved to be highly successful and significantly improved the results of all configurations with respect to both the considered optimization metrics and the optimized quantities of interest. Thus, optimal candidates with $w_{ADP}$ improved by 10 to 17\% and optimal candidates with $w_{OP}$ improved by 25 to 30\% were obtained. Trade-off optimal candidates improving both quantities were also obtained with simultaneous improvements up to 10\% for $w_{ADP}$  and 22\% for $w_{OP}$. \par

Considering that RANS solvers cannot capture the wealth of physical phenomena characteristic of turbomachinery flows, in the future, we are planning on assessing the robustness of our proposed strategies in case of RANS/LES multi-fidelity optimization.

\section*{Acknowledgement}
This work was supported by European Union funding under grant number 101138080 (Project Sci-Fi-Turbo). Views and opinions expressed are however those of the authors only and do not necessarily reflect those of the European Union. Neither the European Union nor the granting authority can be held responsible for them. In addition, the authors gratefully acknowledge Frédéric Alauzet from INRIA Saclay for giving us access to \textit{wolf}  and \textit{feflo.a}, and for his help in using them.

\appendix
\section{POD for shape parameterization data reduction}
\label{app:pod}
This section explains how to reduce the dimension of an FFD shape parameterization with POD \cite{matar_camille_analysis_2024, cinquegrana_efficient_2017}.  \\\par

Starting off from a baseline geometry made of $N_p$ coordinates $(x_b, y_b) \in \mathbb{R}^{N_p \times 2}$, $N_s$ deformation samples are produced with a latin hypercube sampling (LHS) and used to build a database matrix:
\begin{equation*}
    S = \left[y_1, \ldots, y_{N_s}\right] \in \mathbb{R}^{N_p \times N_s}.
\end{equation*}
Then, the fluctuation matrix $F$ is computed from $S$ and $\bar{S}$ as:
\begin{equation*}
    F=S - \bar{S} \in \mathbb{R}^{N_p \times N_s}, \text{ with } \bar{S} = \frac{1}{N_p} \sum_{i=1}^{N_p} y_i.
\end{equation*}
Next, the correlation matrix $\mathbb{C}$  of shape $\mathbb{R}^{N_s \times N_s}$ is computed from:
\begin{equation*}
    \mathbb{C} =F^t \, F,
\end{equation*}
and yields the following eigenvalue problem:
\begin{equation*}
    \mathbb{C}\, V = \Lambda \, V.
\end{equation*}
From the solution $V$, the eigenfunctions matrix $\Phi$ is simply given by:
\begin{equation*}
    \Phi = F\, V \in \mathbb{R}^{N_p \times N_s}.    
\end{equation*}
The reduced designs $S^\star$ are finally given by:
\begin{equation*}
    S^\star = \bar{S} + \Phi^\star\, V^{-1}{}^\star,
\end{equation*}
with $\Phi^\star \in \mathbb{R}^{N_p \times d^\star}$ and $V^{-1}{}^\star \in \mathbb{R}^{d^\star \times N_s}$. At that point, $V^{-1}{}^\star$ is the matrix:
\begin{equation*}
    V^{-1}{}^\star = \left[\alpha_1, \ldots, \alpha_{N_s}\right] \in \mathbb{R}^{d^\star \times N_s},
\end{equation*}
where each modal coefficient column $\alpha_i$ is associated to a reduced profile:
\begin{equation}
    y_i^\star = y_i + \epsilon = \bar{S} + \phi^\star\, \alpha_i,
\end{equation}
and $\epsilon$ denotes reconstruction error. From this last equation, it is hence possible to compute the reduced parameterization $\alpha$ corresponding to any given profile $y$:
\begin{equation}
	\alpha \simeq (\phi^\star{}^t\phi^\star)^{-1}\phi^\star{}^t\, y - (\phi^\star{}^t\phi^\star)^{-1} \phi^\star{}^t\, \bar{S}.
\end{equation}

\begin{figure}[!htbp]
  \centering
  \includegraphics[width=0.5\linewidth]{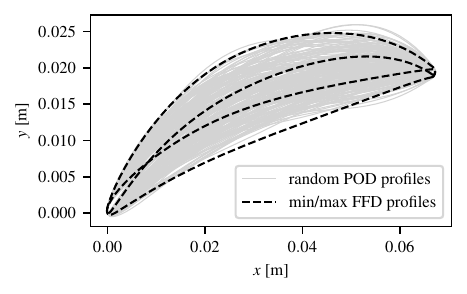}
  \caption{Comparison between the FFD extreme profiles and 100 random reduced profiles sampled via LHS.}
  \label{fig:lrn-cascade-pod-lhs}
\end{figure}

After reduction, the displacement boundaries can be conserved to some extent by sampling coefficients within the extrema obtained in $V^{-1}{}^\star$, but the extreme geometries computed with FFD are not exactly conserved. This is illustrated in \autoref{fig:lrn-cascade-pod-lhs} where some reduced designs violate the extreme profiles. This aspect can be particularly problematic if the meshing and simulation robustness are limited to strict deformation ranges but it is not the case in our context.

\section{Optimal candidates Mach fields}
\label{app:opt}
\begin{figure}[!htp]
\centering
\includegraphics[width=.32\textwidth]{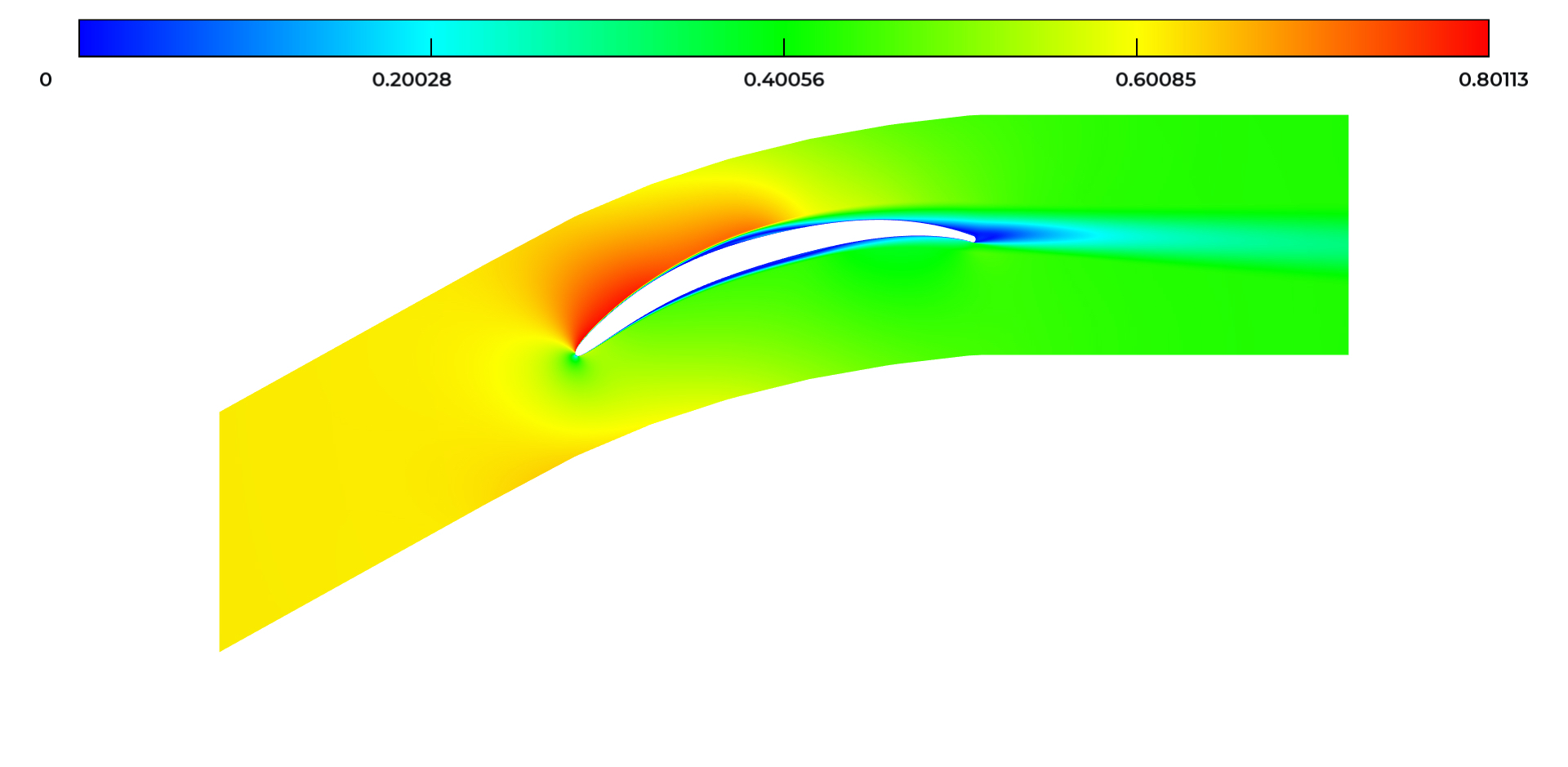}\hfill
\includegraphics[width=.32\textwidth]{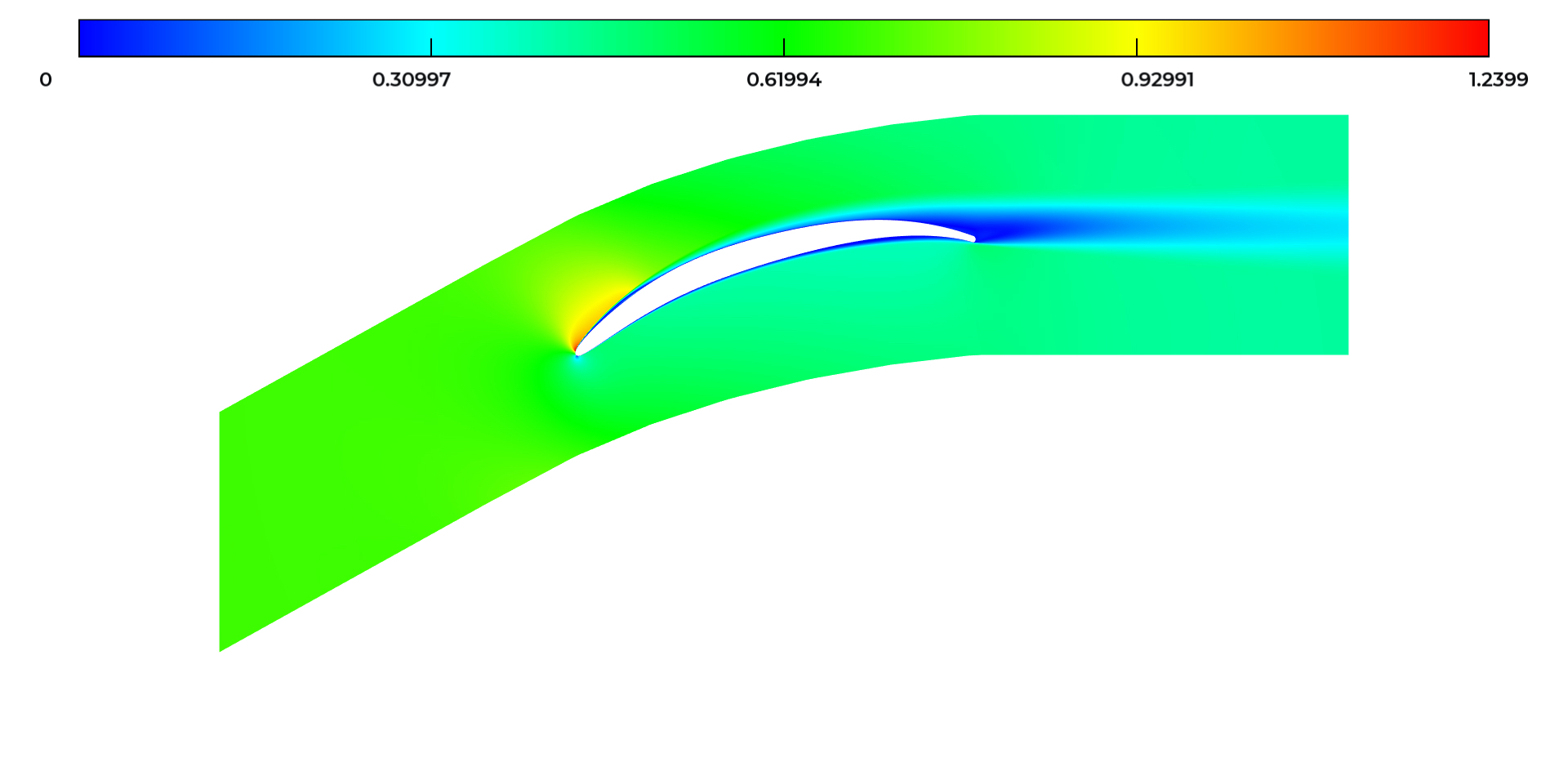}\hfill
\includegraphics[width=.32\textwidth]{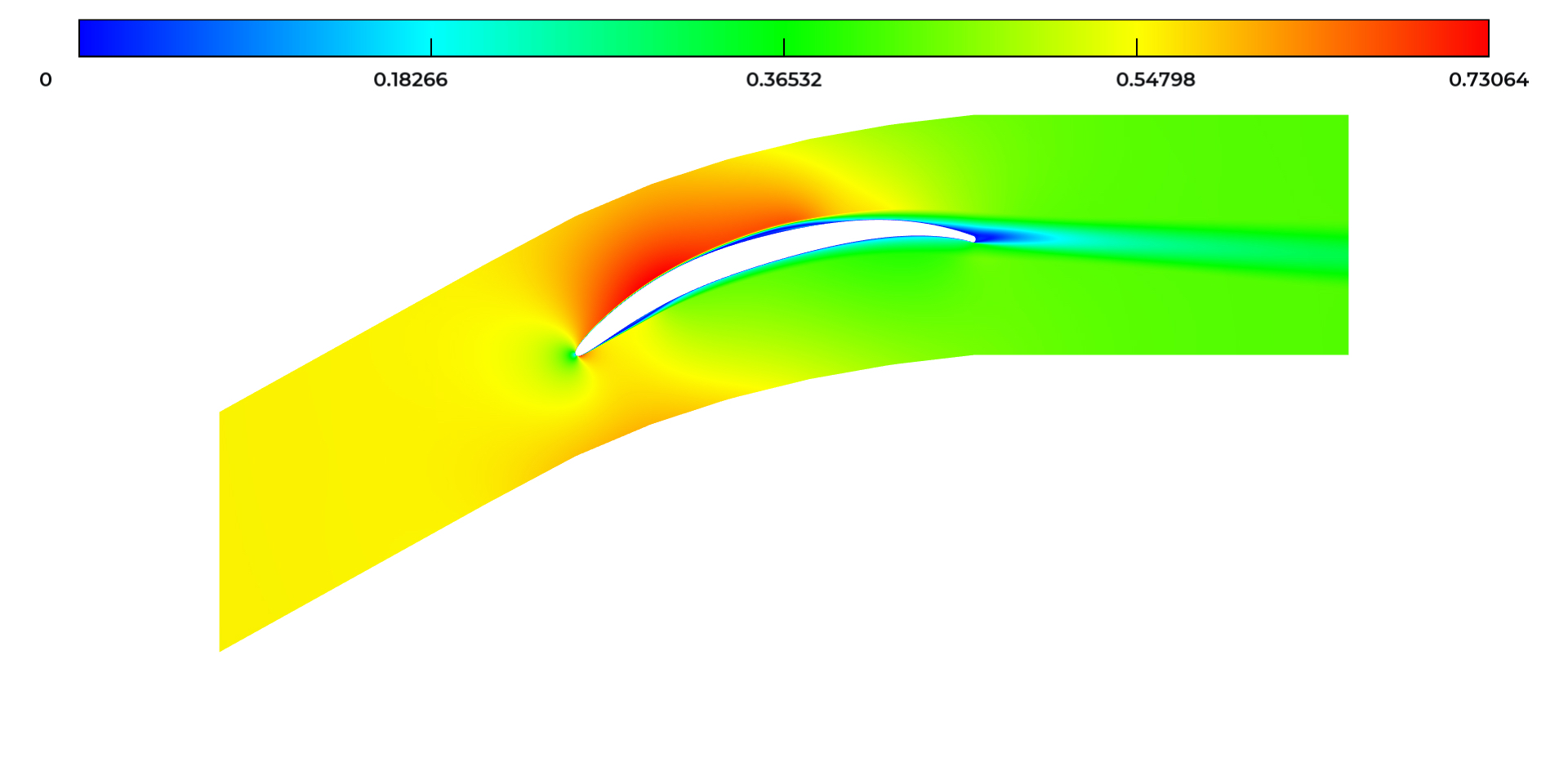}
\caption{ADP (left), OP1 (center), OP2 (right) Mach fields of the candidate with minimal $w_{OP}$. Loss coefficients values are: $w_{ADP}=0.04274$, $w_{OP1}=0.04617$ and $w_{OP2}=0.04049$.}
\label{fig:opt-op}
\end{figure}

\begin{figure}[!htp]
\centering
\includegraphics[width=.32\textwidth]{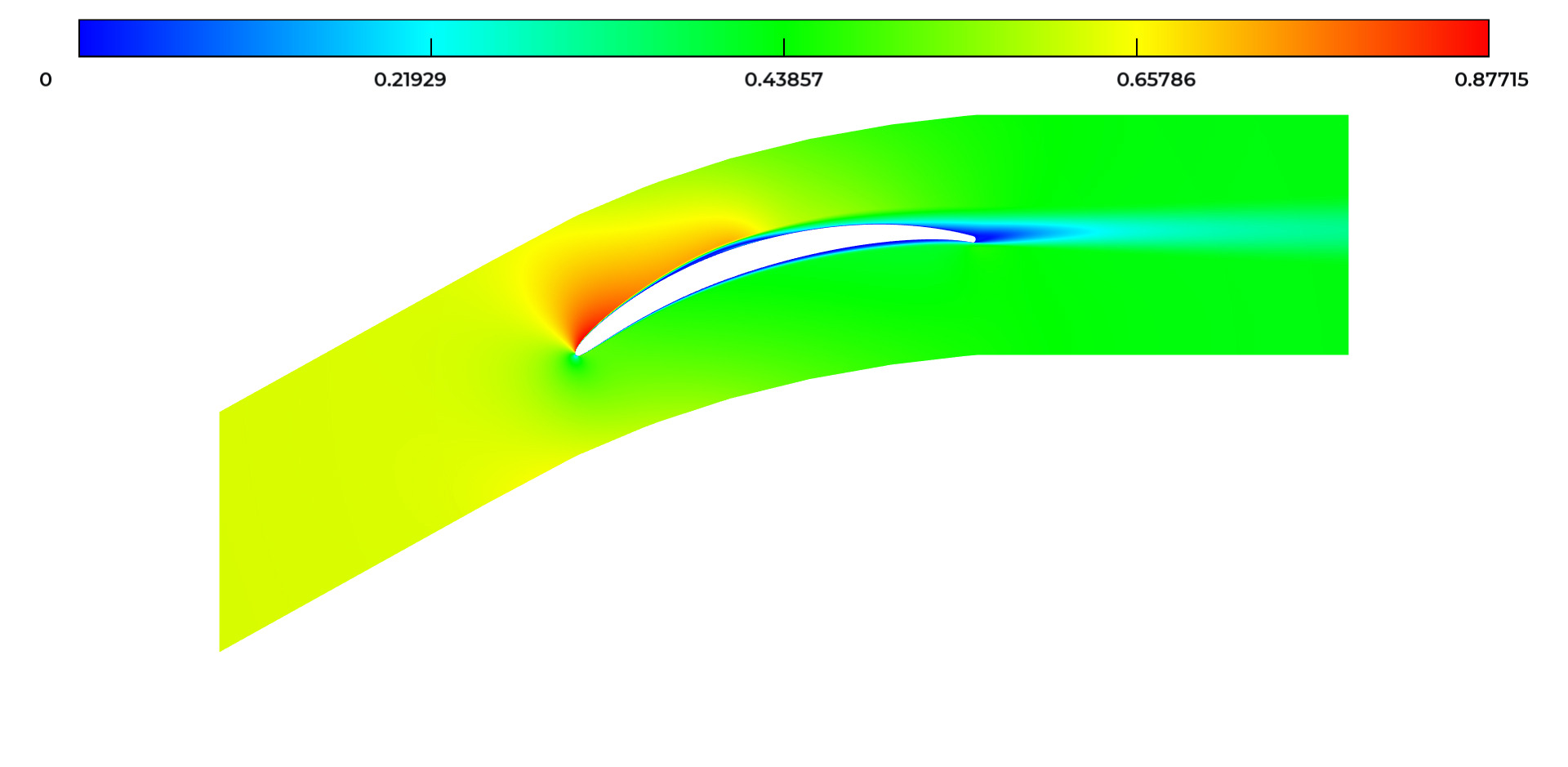}\hfill
\includegraphics[width=.32\textwidth]{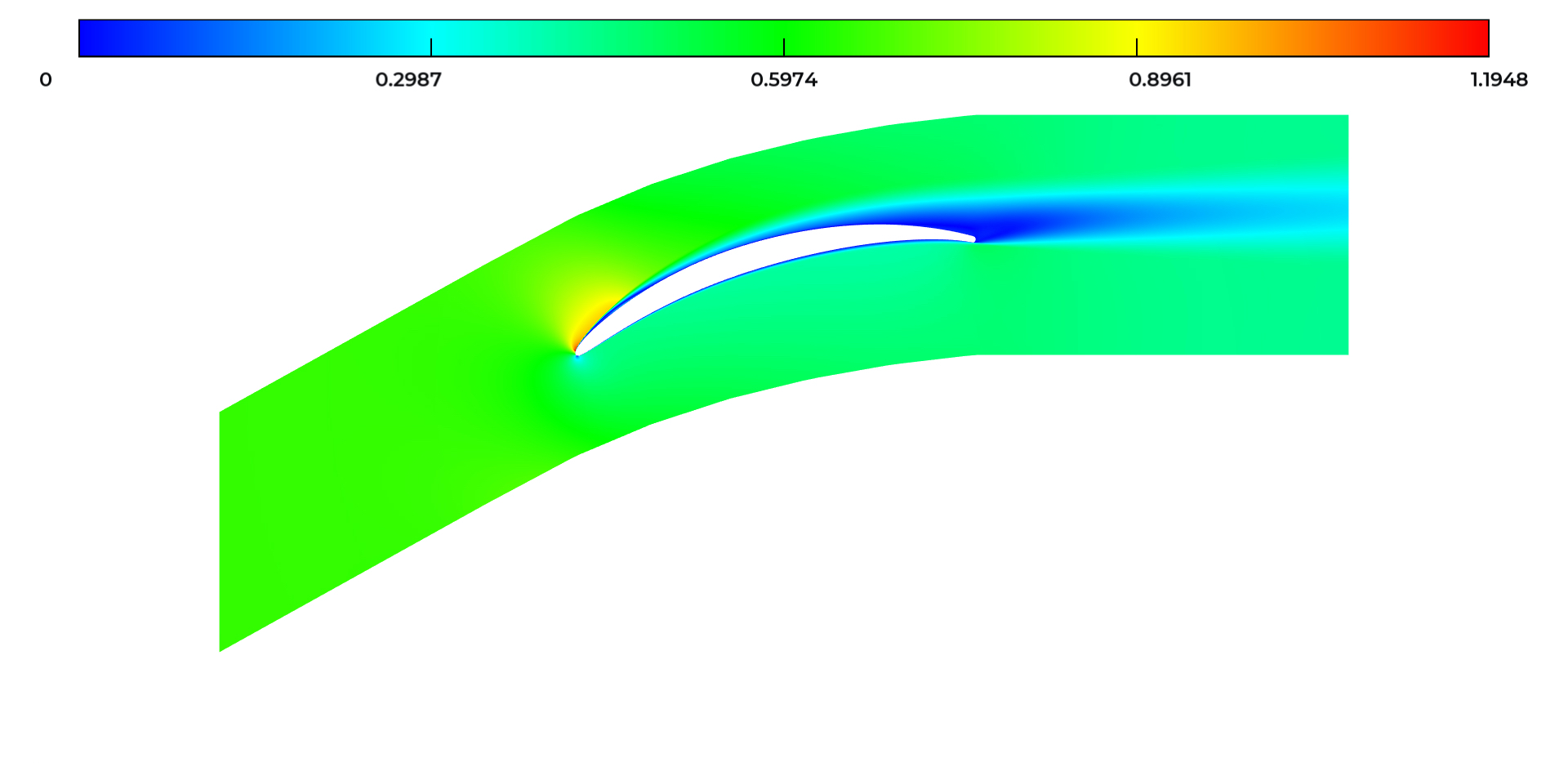}\hfill
\includegraphics[width=.32\textwidth]{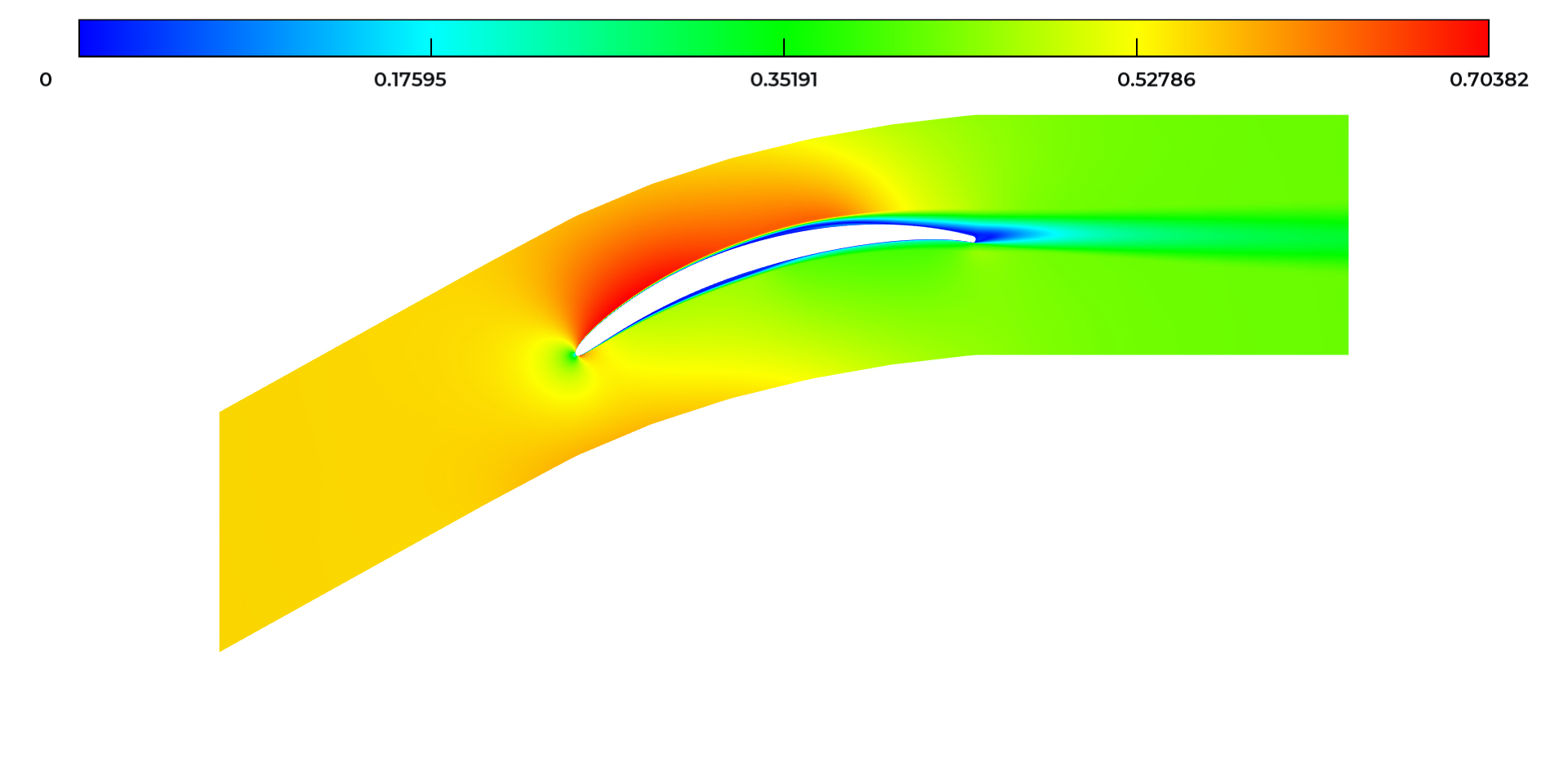}
\caption{ADP (left), OP1 (center), OP2 (right) Mach fields of the candidate with the best compromise. Loss coefficients values are: $w_{ADP}=0.03354$, $w_{OP1}=0.06473$ and $w_{OP2}=0.04286$.}
\label{fig:opt}
\end{figure}

This section presents the three operating points Mach fields for the best candidates selected on the reference Pareto front. \author{fig:opt-adp} for the candidate with the lowest $w_{ADP}$, \autoref{fig:opt-op} for the candidate with the lowest $w_{OP}$ and \autoref{fig:opt} for the candidate at the center of the Pareto front which offers the best compromise between each objective. \\ \par

\begin{figure}[!htp]
\centering
\includegraphics[width=.32\textwidth]{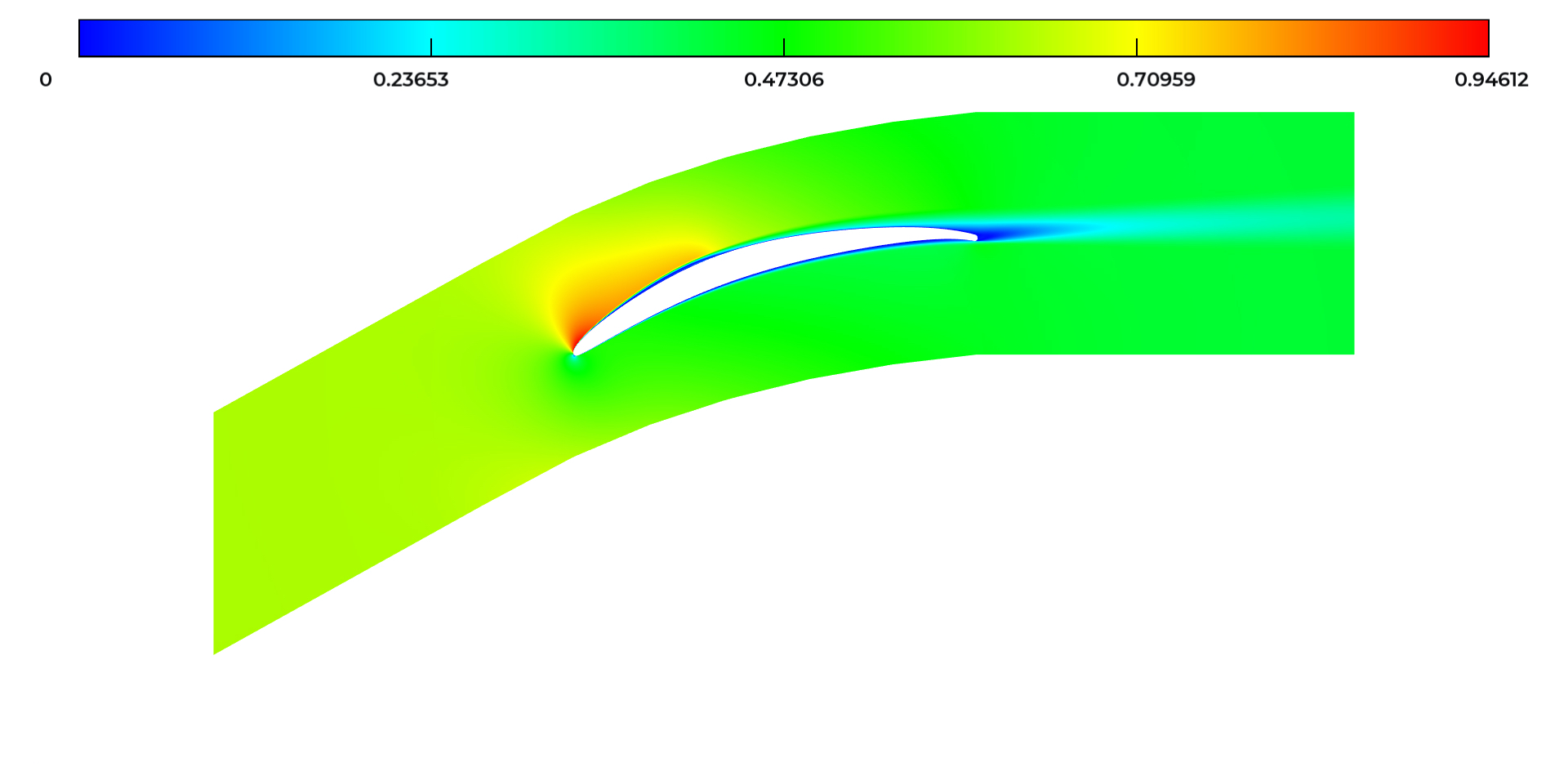}\hfill
\includegraphics[width=.32\textwidth]{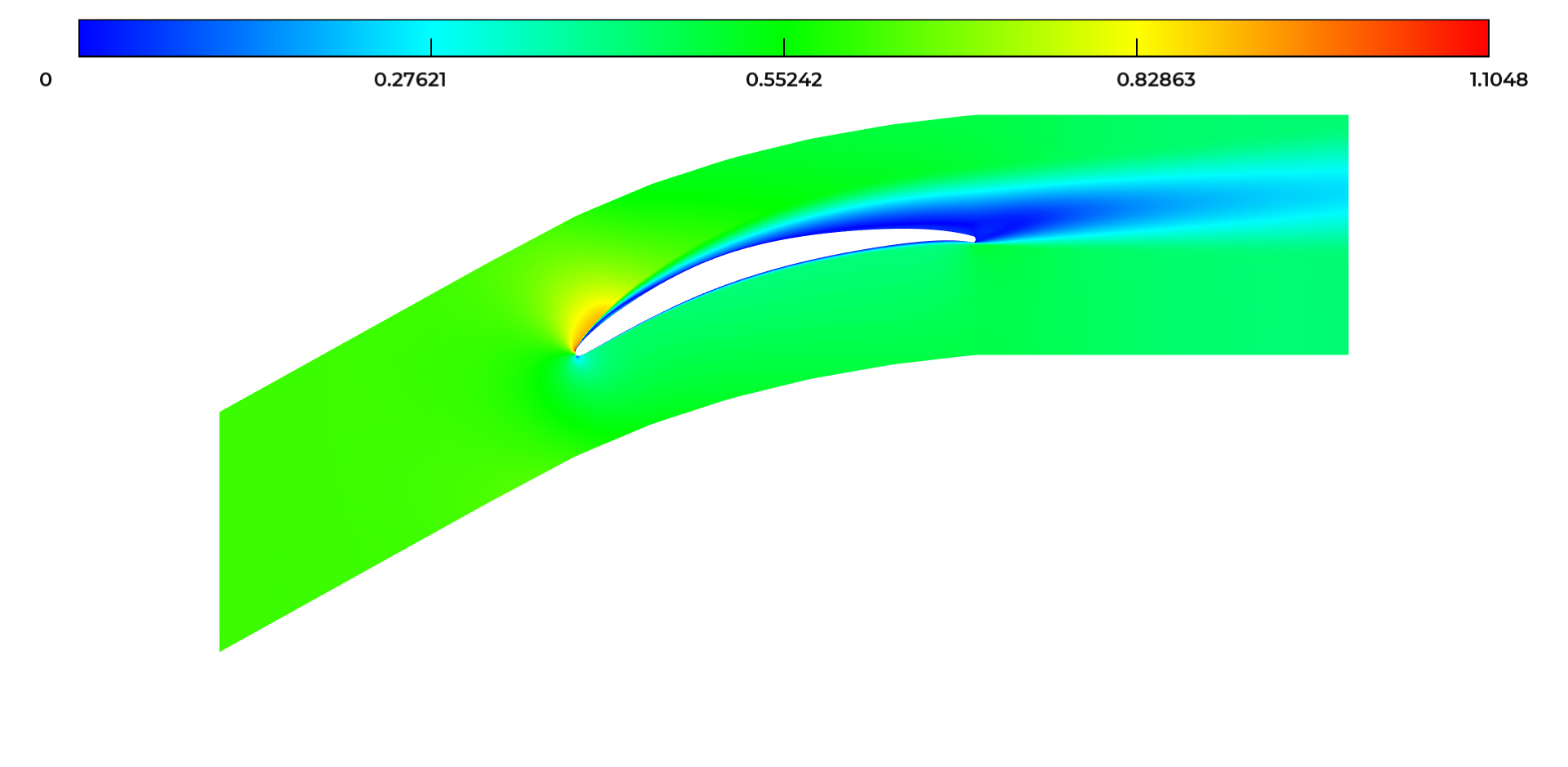}\hfill
\includegraphics[width=.32\textwidth]{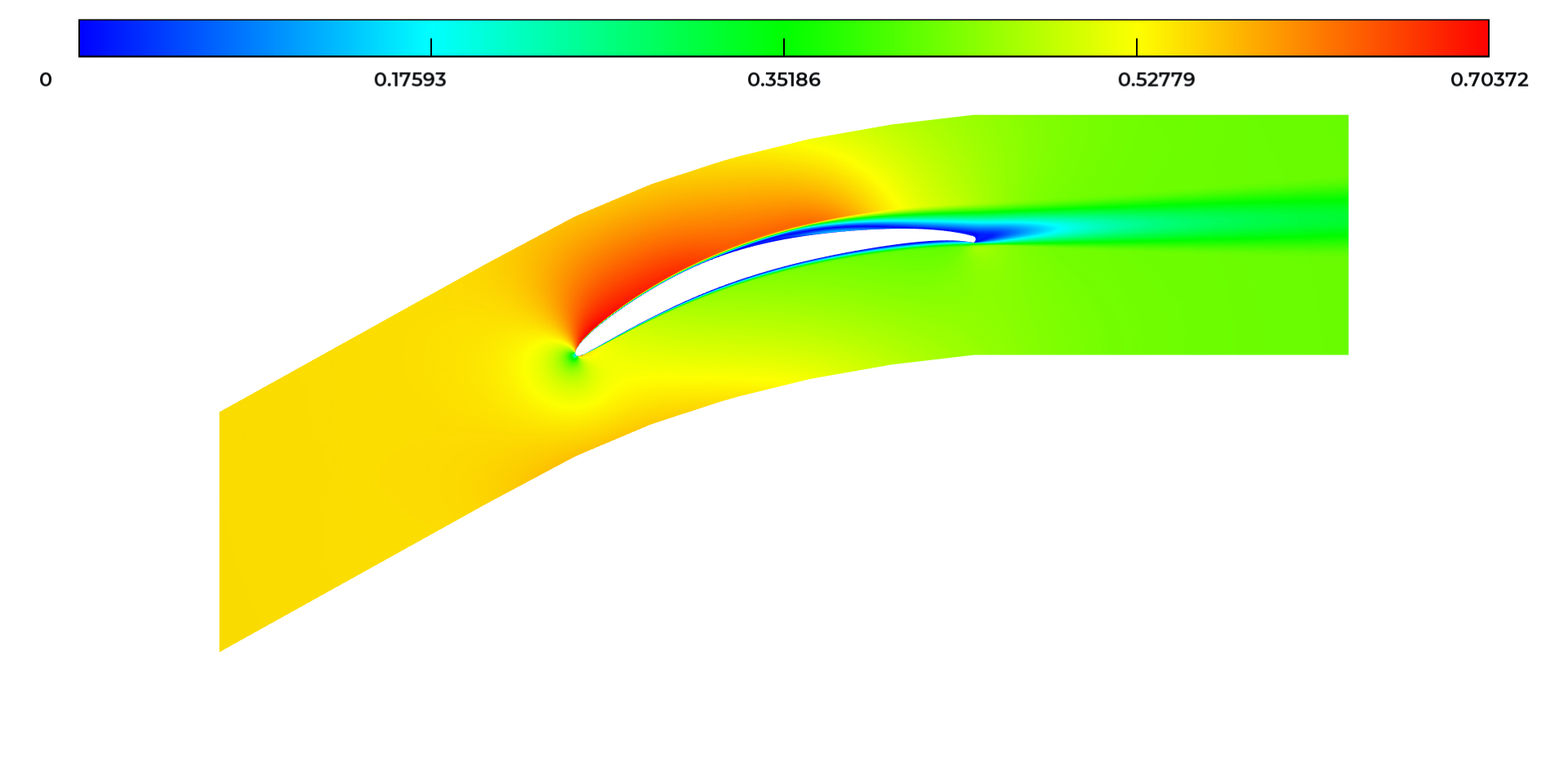}
\caption{ADP (left), OP1 (center), OP2 (right) Mach fields of the candidate with minimal $w_{ADP}$. Loss coefficients values are: $w_{ADP}=0.02925$, $w_{OP1}=0.08938$ and $w_{OP2}=0.04789$.}
\label{fig:opt-adp}
\end{figure}

\newpage
\bibliographystyle{elsarticle-num}
\bibliography{references}

\end{document}